\documentclass[preprint,authoryear,5p,twocolumn]{elsarticle}
\usepackage{graphicx,epsfig}
\usepackage{dcolumn}
\usepackage{bm}
\usepackage{psfrag}
\usepackage{plain}
\usepackage{subfig}
\usepackage{tabularx}
\usepackage[latin1]{inputenc}
\usepackage{amsmath}
\usepackage{amssymb}
\usepackage{graphics}
\usepackage{graphicx}
\usepackage{epsfig}
\usepackage{amssymb}
\usepackage{amsthm}
\usepackage{multirow}
\usepackage[usenames,dvipsnames]{color}
\usepackage[svgnames]{xcolor}

\journal{Aeolian Research}

\begin{document}

\begin{frontmatter}


\title{Origins of barchan dune asymmetry: insights from numerical simulations}


\author[mss]{Eric J. R. Parteli\corref{ejrp}}\ead{Eric.Parteli@cbi.uni-erlangen.de}\author[marum]{Orencio Dur\'an}\author[lpi,dublin]{Mary C. Bourke}\author[israel]{Haim Tsoar}\author[mss]{Thorsten P\"oschel}\author[eth,df]{Hans J. Herrmann}
\address[mss]{Institut f\"ur Multiscale Simulation, Universit\"at Erlangen-N\"urnberg, N\"agelsbachstr.~49b, 91052 Erlangen, Germany}
\address[marum]{MARUM --- Center for Marine Environmental Sciences, University of Bremen, 
28359 Bremen, Germany}
\address[lpi]{Planetary Science Institute, 1700 E Ft Lowell, $\#106$, Tucson AZ, USA}
\address[dublin]{Department of Geography, Trinity College Dublin, Ireland}
\address[israel]{Dep.~Geogr.~Environ.~Development, Ben-Gurion Univ.~Negev, Beer Sheva 84105, Israel}
\address[eth]{Institut f\"ur Baustoffe IfB, ETH H\"onggerberg, HIF E 12, CH-8093, Z\"urich, Switzerland.}
\address[df]{Departamento de F\'{\i}sica, Universidade Federal do Cear\'a - 60455-760, Fortaleza, Cear\'a, Brazil}


\begin{abstract}
Barchan dunes --- crescent-shaped dunes that form in areas of unidirectional winds and low sand availability --- commonly display an asymmetric shape, with one limb extended downwind. Several factors have been identified as potential causes for barchan dune asymmetry on Earth and Mars: asymmetric bimodal wind regime, topography, influx asymmetry and dune collision. However, the dynamics and potential range of barchan morphologies emerging under each specific scenario that leads to dune asymmetry are far from being understood. In the present work, we use dune modeling in order to investigate the formation and evolution of asymmetric barchans. We find that a bimodal wind regime causes limb extension when the divergence angle between primary and secondary winds is larger than $90^{\circ}$, whereas the extended limb evolves into a seif dune if the ratio between secondary and primary transport rates is larger than $25\%$. Calculations of dune formation on an inclined surface under constant wind direction also lead to barchan asymmetry, however no seif dune is obtained from surface tilting alone. Asymmetric barchans migrating along a tilted surface move laterally, with transverse migration velocity proportional to the slope of the terrain. Limb elongation induced by topography can occur when a barchan crosses a topographic rise. Furthermore, transient asymmetric barchan shapes with extended limb also emerge during collisions between dunes or due to an asymmetric influx. Our findings can be useful for making quantitative inference on local wind regimes or spatial heterogeneities in transport conditions of planetary dune fields hosting asymmetric barchans.
\end{abstract}


\end{frontmatter}




\section{\label{sec:introduction}Introduction}
\begin{figure}[!t] 
\begin{center} 
\vspace{0.1cm}
\includegraphics[width=0.85\columnwidth]{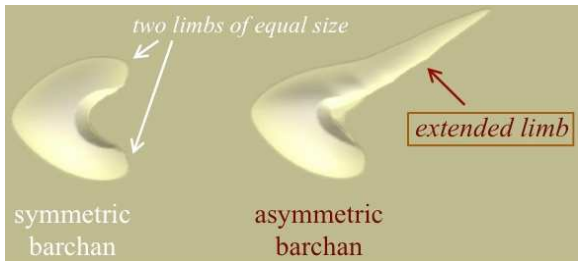} 
\caption{Schematic diagram of a symmetric barchan with two limbs of equal size (left) and of an asymmetric barchan with an extended limb (right).}
\label{fig:sketch_barchan_asymmetry} 
\end{center} 
\end{figure}

The classical symmetric shape of barchan dunes is far from being prevalent in nature. A wide variety of barchan morphologies on Earth and Mars are asymmetric, with one extended limb (cf.~Fig.~\ref{fig:sketch_barchan_asymmetry}) \citep{Tsoar_1984,Bourke_2010}. Since the pioneering works by \cite{Bagnold_1941}, various conceptual models have been proposed in order to explain the existing asymmetric dune morphologies. Dune asymmetry has been most predominantly attributed to asymmetric bimodal winds \citep{Bagnold_1941,Tsoar_1984}, topography \citep{Finkel_1959,Long_and_Sharp_1964}, dune collisions \citep{Close_Arceduc_1969,Hersen_and_Douady_2005} or asymmetric sand influx \citep{Rim_1958}. Indeed, understanding dune asymmetry constitutes an important issue in planetary science, as asymmetric dunes could potentially serve as a proxy for local wind regimes or variations in sand supply or topography. However, the significance of the different causes of barchan asymmetry is still poorly understood, whereas the potential range of dune morphologies resulting in each case remains unknown. Although some insights into the dynamics of asymmetric dunes could be gained from field monitoring within a time span of a few years \citep{Bourke_2010}, assessment of dune shape evolution is a difficult task owing to the long timescales involved in dune processes. Moreoever, diverse asymmetric dune morphologies \citep{Bourke_et_al_2004,Bourke_and_Goudie_2009} might be the outcome of different factors in concurrent action, which poses a further challenge in the investigation of the developmental stages of asymmetric dunes through field observations. 

Numerical modeling --- which has become indispensable in the investigation of aeolian and dune processes on Earth and other planetary bodies \citep{Bourke_et_al_2010,Kok_et_al_2012} --- could provide a helpful tool in the study of barchan dune asymmetry. In order to shed light on the factors competing for the appearance of various asymmetric dune shapes, a minimal model that accounts for a mathematical description of the wind over dunes, as well as for the equations for grain transport and landscape evolution is required. Such a model has been developed in the course of the last decade \citep{Sauermann_et_al_2001,Kroy_et_al_2002,Duran_and_Herrmann_2006a,Duran_et_al_2010}. This model combines an analytical model for the average turbulent flow over the sand landscape with a continuum model for sand transport. The model has proven to reproduce the shape of different types of dunes with good quantitative agreement with measurements \citep{Sauermann_et_al_2003,Duran_and_Herrmann_2006a,Duran_and_Herrmann_2006b,Parteli_and_Herrmann_2007}. The model has been further extended in order to model longitudinal seif dunes and diverse unusual Martian dune forms under bimodal wind regimes \citep{Parteli_et_al_2009}.

In the present work, we use the dune model to investigate the role of different asymmetry causes for the diversity of asymmetric barchan shapes reported previously from observations of dune fields on Earth and Mars \citep{Bourke_2010}. We calculate the evolution of a symmetric barchan dune under the separate action of the potential causes for dune asymmetry.

This paper is organized as follows. In Section \ref{sec:model} the dune model is described. In Sections \ref{sec:bimodal_winds}-\ref{sec:dune_collision} we present the results obtained from calculations of barchan dune asymmetry due to bimodal wind regimes, topography, asymmetric sand supply and dune collision, respectively. Moreover, in Section \ref{sec:discussion} we compare the results of our simulations with the shapes of real asymmetric barchans and discuss on the application of the model outcomes to the research of extraterrestrial dunes. Finally, conclusions are presented in Section \ref{sec:conclusions}.

\section{\label{sec:model}Model description}

The model used in the calculations of the present work consists of a set of mathematical equations that compute: the average turbulent wind field (the surface shear stress (${\boldsymbol{\tau}}$) over the topography; the mass flux (${\boldsymbol{q}}$) of saltating particles due to the shear stress, and the time evolution of the surface resulting from particle transport \citep{Sauermann_et_al_2001,Kroy_et_al_2002,Duran_and_Herrmann_2006a}. In this Section, we present a brief description of the model equations and of the calculation procedure. 

\subsection{\label{sec:wind_field}Wind field}

The first step of the model calculations consists of describing quantitatively the average three-dimensional turbulent wind field over the dune. The average shear stress field (${\boldsymbol{{\tau}}}$) is calculated through solving a set of analytical equations developed by \cite{Weng_et_al_1991}. 

In the absence of dunes, the wind velocity ${\boldsymbol{v}}(z)$ within the atmospheric boundary layer increases logarithmically with the height ($z$) above the flat ground. That is, 
\begin{equation}
{\boldsymbol{v}}(z) = [{\boldsymbol{u}}_{{\ast}0}/{\kappa}]\,{\mbox{ln}}(z/z_0),
\end{equation}
where $\kappa = 0.4$ is the von K\'arm\'an constant and ${\boldsymbol{u}}_{{\ast}0}$ is the wind shear velocity, which is used to define the (undisturbed) shear stress ${\boldsymbol{{\tau}}}_0 = {\rho}_{\mathrm{fluid}}|{\boldsymbol{u}}_{{\ast}0}|{\boldsymbol{u}}_{{\ast}0}$, with ${\rho}_{\mathrm{fluid}}$ standing for the air density \citep{Bagnold_1941,Sullivan_et_al_2000}. Furthermore, $z_0$ is the surface roughness, which scales with the average grain size ($d$) composing the sand bed. We take $z_0 \approx d/20$ based on a recent theoretical work on modeling saturated sand flux \citep{Duran_and_Herrmann_2006a}.

A smooth hill or dune introduces a perturbation in the wind field. The Fourier-transformed longitudinal and transverse components of the shear stress perturbation $({\boldsymbol{\hat{\tau}}}$) due to the local topography are computed using following equations,
\begin{equation}
{\tilde{{{\hat{\tau}}}}}_x = {\frac{{{\tilde{h}}_{\mathrm{s}}}k_x^2}{|\vec{k}|}}{\frac{2}{U^2(l)}}{\left\{{-1 + {\left({2{\ln{{\frac{l}{z_0^{\prime}}}}} + {\frac{{|{\vec{k}}|}^2}{k_x^2}}}\right)}{\sigma}{\frac{K_1(2{\sigma})}{K_0(2{\sigma})}}}\right\}}, \label{eq:wind_A}
\end{equation}
\begin{equation}
{\tilde{{{\hat{\tau}}}}}_y = {\frac{{{\tilde{h}}_{\mathrm{s}}}k_xk_y}{|\vec{k}|}}{\frac{2}{U^2(l)}}2{\sqrt{2}}{\sigma}K_1(2{\sqrt{2}}{\sigma}), \label{eq:wind_B}
\end{equation}
where $x$ and $y$ are parallel, respectively, perpendicular to the wind direction, $\sigma = \sqrt{{\mbox{i}}Lk_xz_0/l}$, $k_x$ and $k_y$ are the components of the wave vector $\vec{k}$, i.e. the coordinates in Fourier space, $K_0$ and $K_1$ are modified Bessel functions and ${\tilde{h}}_{\mathrm{s}}$ is the Fourier transform of the height profile; $U$ is the vertical velocity profile which is suitably non-dimensionalized, $l$ is the depth of the inner layer of the flow and $L$ is a typical length scale of the hill or dune and is given by $1/4$ the mean wavelength of the Fourier representation of the height profile. In the presence of the saltating grains, the roughness of the dune's surface increases to an apparent value $z_0^{\prime}$, the aerodynamic roughness \citep{Bagnold_1941}. The wind velocity over the dune is calculated using $z_0^{\prime} = 1$ mm, a value based on experimental observations \citep{Bagnold_1941,Andreotti_2004}. The surface shear stress field is obtained, then, with the equation, 
\begin{equation}
{\boldsymbol{\tau}} = |{\boldsymbol{{\tau}}}_0|{({{{\boldsymbol{{\tau}}}_0}/|{\boldsymbol{{\tau}}}_0| + {\boldsymbol{\hat{\tau}}}})}, \label{eq:shear_stress}
\end{equation}
where ${\boldsymbol{{\tau}}}_0$ is the undisturbed shear stress over the flat ground.

Flow separation at the dune brink due to the strong local curvature of the surface gives rise to a zone of recirculating flow \citep{Walker_and_Nickling_2002,Herrmann_et_al_2005,Araujo_et_al_2013}, which cannot be described by the analytical model \citep{Weng_et_al_1991}. This problem is overcome introducing the so-called ``separation bubble'': for each longitudinal slice of the dune, a separation streamline connecting the brink to the reattachment point is introduced at the lee. Each separation streamline is fitted to a third-order polynomial, the parameters of which are determined as described in detail by \cite{Kroy_et_al_2002}. The wind model \citep{Weng_et_al_1991} is then solved for the smooth ``envelope'' that comprises the separation bubble and the dune surface \citep{Kroy_et_al_2002}. Thereafter, the shear stress in the recirculating zone within the separation bubble is set as zero, since the net transport within the bubble essentially vanishes.
\subsection{Sand flux}

Next, the mass flux of particles in saltation --- which is the dominant transport mode and consists of grains hoping in ballistic trajectories and ejecting new particles upon collision with the ground --- is computed. The saltation cloud is regarded as a thin, fluid-like layer that can exchange sand with the immobile sand bed \citep{Sauermann_et_al_2001}. 

When the wind shear stress exceeds a minimal threshold and saltation begins, the sand flux first grows exponentially due to the multiplicative process inherent to the saltation process. However, since saltating grains accelerate at cost of aeolian momentum, the flux cannot increase beyond a maximal value, the so-called saturated sand flux \citep{Bagnold_1941,Sauermann_et_al_2001,Almeida_et_al_2008,Duran_et_al_2011,Paehtz_et_al_2012} which is reached after a saturation transient where the air shear stress within the saltation cloud equals the minimal value for sustained saltation, i.e. the impact threshold, ${\tau}_{\mathrm{t}} = {\rho}_{\mathrm{fluid}}u_{{\ast}{\mathrm{t}}}^2$ \citep{Bagnold_1941}, defined in terms of the impact threshold shear velocity $u_{{\ast}{\mathrm{t}}}$. In the continuum model, the derivation of the three-dimensional equations for the sand flux is performed by explicitly accounting for the decrease in aeolian shear stress due to the growth of the number of particles in saltation (the  ``feedback effect'' \citep{Owen_1964}), which leads to the saturation transient of the flux as described above. The following equation is obtained for the sand flux (${\boldsymbol{q}}$) over the terrain,
\begin{equation}
{\boldsymbol{\nabla}}{\cdot}{\boldsymbol{q}} = (1 - |{\boldsymbol{q}}|/q_{\mathrm{s}})|{\boldsymbol{q}}|/{\ell}_{\mathrm{s}}, \label{eq:sand_flux}
\end{equation}
 where $q_{\mathrm{s}} = [2{\alpha}|{\boldsymbol{v_{\mathrm{s}}}}|/g]({{\tau} - {\tau}_{\mathrm{t}}})$ is the saturated flux; ${\ell}_{\mathrm{s}} = [2{\alpha}{|{\boldsymbol{v_{\mathrm{s}}}}|^2}/g{\gamma}]{\tau}_{\mathrm{t}}{({{\tau} - {\tau}_{\mathrm{t}}})}^{-1}$ is the characteristic length of flux saturation; $g$ is gravity; $\alpha \approx 0.4$ and $\gamma \approx 0.2$ are empirically determined model parameters \citep{Sauermann_et_al_2001,Duran_and_Herrmann_2006a}, and the steady-state velocity of the particles in saltation (${\boldsymbol{v_{\mathrm{s}}}}$) is calculated numerically from the balance between aeolian drag, gravitational and bed friction forces on the particles \citep{Duran_et_al_2010}. The outcome of the calculation is a two-dimensional field (${\boldsymbol{q}}(x,y$)) that gives the height-integrated, average mass flux of saltating particles over the terrain. 

\subsection{Surface evolution}

Finally, mass conservation is used in order to compute the evolution of the local height ($h(x,y)$) through the equation,
\begin{equation}
{\rho}_{\mathrm{sand}}{\partial}h/{\partial}t = -{\nabla}\cdot{\boldsymbol{q}}, \label{eq:mass_conservation}
\end{equation} 
where ${\rho}_{\mathrm{sand}}$ is the bulk sand density. Eq. (\ref{eq:mass_conservation}) implies that the deposition (erosion) occurs at those places where the flux locally decreases (increases) downwind. 

{\em{Avalanches}} --- wherever the local slope exceeds the angle of repose of the sand (${\theta}_{\mathrm{c}} \approx 34^{\circ}$), the surface is relaxed through avalanches in the direction of the steepest descent. Avalanches are considered to be instantaneous as their time-scale can be regarded as negligible compared to the time-scale of the surface evolution due to aeolian transport. The flux of avalanches along the slip-face is given by,
\begin{equation}
{\boldsymbol{q}}_{\mathrm{aval}} = k{\left[{ {\mbox{tanh}}({{\nabla}h}) - {\mbox{tanh}}({\theta}_{\mathrm{dyn}}) }\right]}{\frac{{\nabla}h}{|{\nabla}h|}}, \label{eq:avalanche_flux}
\end{equation}
where $k = 0.9$ is a parameter and ${\theta}_{\mathrm{dyn}} = 33^{\circ}$ is the so-called ``dynamic'' angle of repose, which characterizes the surface after relaxation \citep{Duran_et_al_2010}. The update in the local height is obtained by solving Eq. (\ref{eq:mass_conservation}) using the flux due to avalanches given by Eq. (\ref{eq:avalanche_flux}). The calculation is repeated until the local slope is below ${\theta}_{\mathrm{dyn}}$.

\subsection{Description of the calculation procedure}

Summarizing, the model consists of iteratively performing the following calculations:
\begin{enumerate}
\item the average shear stress (${\boldsymbol{\tau}}$) over the surface is computed using Eqs. (\ref{eq:wind_A}), (\ref{eq:wind_B}) and (\ref{eq:shear_stress});
\item next, the height-integrated average mass flux ${\boldsymbol{q}}$ over the terrain is calculated by solving Eq. (\ref{eq:sand_flux});
\item the change in the local surface is computed with Eq. (\ref{eq:mass_conservation}); wherever the local inclination is larger than $34^{\circ}$, the flux due to avalanches is calculated using Eq. (\ref{eq:avalanche_flux}) and the topography updated again using Eq. (\ref{eq:mass_conservation}).
\end{enumerate}
The initial surface is a smooth hill of Gaussian shape, which is subjected to a constant upwind shear stress of value ${\tau}_0$ --- in the following, we call the upwind shear stress simply ${\tau}$. 

Calculations are performed with open boundaries and an influx $q_{\mathrm{in}}$ at the inlet, which is a fraction of the saturated flux, $q_{\mathrm{s}}$. Indeed, field observations show that transport in interdune areas is typically in the undersaturated regime and is dependent upon several factors, predominantly terrain type and the size and shape of dunes upwind \citep{Fryberger_et_al_1984}. We consider that the interdune flux in a field of barchans migrating on top of bedrock is dictated by the average output flux of the dunes in the field (that is, the flux of sediment being released from the barchans' limb). Since the output flux of barchan dunes produced with the model is about $20\%$ of the saturated flux \citep{Duran_et_al_2010}, we choose $q_{\mathrm{in}}/q_{\mathrm{s}} = 0.2$. We note that this choice for $q_{\mathrm{in}}/q_{\mathrm{s}}$ is fairly consistent with measurements of drift rates by \cite{Fryberger_et_al_1984} in the Jafurah sand sea in Saudi Arabia. The authors reported monthly-averaged drift rates downwind of a mature barchan varying within a broad range between $0$ and about $67\%$ of the value on top of a sand sheet (where the flux is saturated), with annual mean about $20\%$ (cf.~data for trap $\#6$ in Table 2 by \cite{Fryberger_et_al_1984}). However, we further note that the same authors also reported significantly larger drift rates (about $60\%$) downwind of domes and small dunes with incipient slip-face, while nearly vanishing drift rate values were obtained from measurements in interdune areas consisting of salt-encrusted sandy plains. Therefore, it should be remarked that our choice for $q_{\mathrm{in}}$ should be reasonable for mature barchans separated by bedrock interdune areas. For other situations, like dunes surrounded by sand patches that can serve as sediment source or evolving on a terrain containing vegetation or moisture (thus hindering interdune sediment transport), the value of $q_{\mathrm{in}}$ should be adjusted in order to adequately model the particular physical conditions.

A list of the main relevant parameters of the model with their respective values used in our calculations is displayed in Table \ref{tab:parameters}.

\begin{table*}[!htpb]
\begin{center}
\begin{tabular}{|c|c|c|c|c|c|c|c|c|}
\hline \hline 
$d$ ($\mu$m) & ${\rho}_{\mathrm{p}}$\,(kg$/$m${^3}$) & $\eta$\,(kg/m$\cdot$s) & ${\rho}_{\mathrm{fluid}}$\,(kg$/$m${^3}$) & $g$\,(m$/$s$^2$) & $z_0$\,(${\mu}$m) & $z_0^{\prime}$\,(m) & ${\alpha}$ & $\gamma$ \\
\hline
$250$ & $2650$ & $1.78 \times 10^{-5}$ & $1.225$ & $9.8$ & $1.25$ & $0.001$ & $0.4$ & $0.2$ \\
\hline \hline
\end{tabular}
\caption{Main relevant parameters of the model with their respective values associated with aeolian transport under Earth conditions \citep{Sauermann_et_al_2001,Kroy_et_al_2002,Duran_and_Herrmann_2006a,Duran_et_al_2010}. $d = 250{\mu}$m is the average grain size of Earth dunes \citep{Pye_and_Tsoar_1990} and ${\rho}_{\mathrm{p}} = 2650\,$kg$/$m$^3$ is the density of quartz, while ${\rho}_{\mathrm{fluid}} = 1.225\,$kg$/$m$^3$ and $\eta = 1.78 \times 10^{-5}\,$kg\,m$^{-1}$s$^{-1}$ are the density and dynamic viscosity of the air, respectively. Furthermore, $g = 9.81\,$m$/$s$^2$ is gravity, $z_0 \approx d/20 = 1.25\,{\mu}$m is the surface roughness in the absence of saltation and $z_0^{\prime} = 1\,$mm is the aerodynamic roughness, which accounts for the presence of saltating cloud, while $\alpha = 0.4$ and $\gamma = 0.2$ are empirically determined parameters of the sand flux model \citep{Sauermann_et_al_2001,Duran_and_Herrmann_2006a}. Using these parameters, the threshold shear velocity for sustained transport, $u_{{\ast}{\mathrm{t}}} \approx 0.21\,$m$/$s is calculated with the model by \cite{Iversen_and_White_1982}.}
\label{tab:parameters}
\end{center}
\end{table*}

\section{\label{sec:bimodal_winds}Bimodal wind regimes}

Most attempts to model barchan asymmetry have concentrated on the role of asymmetric bimodal wind regimes for the elongation of one barchan limb \citep{Bagnold_1941,Tsoar_1984} (see also \cite{Bourke_2010}). The bimodal wind is said to be asymmetric when the transport rates of both wind components are not equal.

The first conceptual model was by \cite{Bagnold_1941}. According to this model, a symmetric barchan, originally formed by a gentle wind, becomes asymmetric if a storm wind blows from a secondary direction, making an accute divergence angle with the primary one. The limb exposed to the storm wind elongates as it enters the sand stream of the limb at the opposite side, thus evolving into a longitudinal seif dune \citep{McKee_and_Tibbitts_1964,Tsoar_1982,Tsoar_1983,Bristow_et_al_2000,Rubin_et_al_2008,Tokano_2010}. According to Bagnold's model, the seif dune formed in this manner aligns approximately parallel to the direction of the storm \citep{Bagnold_1941}. This conceptual model was referred to by many authors in the past \citep{Verstappen_1968,Ruhe_1975,Goudie_and_Wilkinson_1977,Mabbutt_1977,Lancaster_1980}, though reports on supporting field examples were scarce \citep{Lancaster_1980}. A different model was proposed later by \cite{Tsoar_1984}: a gentle wind (not a storm wind) blows from the secondary direction, and the limb that elongates is the one opposite to the secondary gentle wind \citep{Tsoar_1984}. Some field observations exist that support this model \citep{Bourke_2010}. 

In our calculations, we simulate a bimodal wind by periodically alternating the orientation of the field between two directions forming a divergence angle, ${\theta}_{\mathrm{w}}$ \citep{Parteli_et_al_2009} --- such as in the turntable experiments of ripples and subaqueous dune formation on a sediment bed under bimodal flow regimes \citep{Rubin_and_Hunter_1987,Rubin_and_Ikeda_1990}. The wind model is solved considering a constant wind blowing over the rotated landscape, while the separation bubble adapts to the wind direction following the rotation of the field \citep{Parteli_et_al_2009}. The {\em{primary}} wind direction, i.e. the one that forms the barchan, has duration $T_{{\mathrm{w}}1}$ and upwind shear stress ${\tau}_1$. The {\em{secondary}} wind direction makes an angle ${\theta}_{\mathrm{w}}$ with the primary wind, and lasts for a time $T_{{\mathrm{w}}2}$ with upwind shear stress ${\tau}_2$. 

\begin{figure}[!t] 
\begin{center} 
\vspace{0.1cm}
\includegraphics[width=0.85\columnwidth]{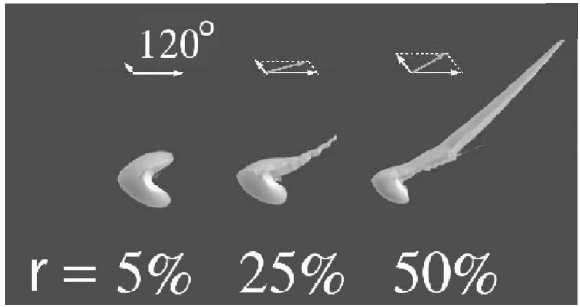} 
\caption{Elongation of one barchan limb due to a bimodal wind regime with obtuse divergence angle. Wind directions are indicated by the arrows (the primary wind, which has duration $T_{{\mathrm{w}}1}$, blows from the left). The barchan shape depends on $r = T_{{\mathrm{w}}2}/T_{{\mathrm{w}}1}$, i.e. the duration $T_{{\mathrm{w}}2}$ of the secondary wind relative to $T_{{\mathrm{w}}1}$. In these simulations, $T_{{\mathrm{w}}1}$ is about $3\%$ of the migration time of the barchan dune, $T_{\mathrm{m}}$. The shapes shown for $r \leq 25\%$ are {\em{stationary}} shapes, whereas for $r > 25\%$, the elongated dune horn increases with time. Simulation snapshots correspond to time $t$ of the order of $5\,T_{\mathrm{m}}$. The barchan obtained in the simulation with $r=5\%$ has length and width of approximately 120~m.}
\label{fig:duran} 
\end{center} 
\end{figure}

We begin our study with an obtuse divergence angle (${\theta}_{\mathrm{w}} = 120^{\circ}$), and consider two models: 

{\em{Model $\#1$}}: both wind directions have the same upwind shear stress (${\tau}_1 = {\tau}_2$), however the primary wind has a longer duration ($T_{{\mathrm{w}}1} > T_{{\mathrm{w}}2}$) --- a scenario reminiscent of the experiments by \cite{Rubin_and_Hunter_1987} with the difference, however, that in the simulations the ground is not covered with sand. Fig.~\ref{fig:duran} shows snapshots of the dune evolution obtained from the calculations. As can be seen from Fig.~\ref{fig:duran}, the barchan develops an asymmetric shape, which depends on the ratio $r \equiv T_{{\mathrm{w}}2}/T_{{\mathrm{w}}1}$. If $r > 25\%$, then the limb at the side opposite to the secondary wind elongates into the {\em{resultant}} wind direction (Fig.~\ref{fig:duran}) to form a longitudinal dune. It is interesting that the longitudinal alignment also appears to be lost in turntable experiments of ripples on a sand bed when the transport rates differ by a factor larger than 4 \citep{Rubin_and_Hunter_1987}. A theoretical model following the concept of maximum gross transport \citep{Rubin_and_Hunter_1987,Rubin_and_Ikeda_1990} could shed light on the origin of the threshold value of $r$ for seif dune elongation. Here we could not find a significant dependence of this threshold on ${\theta}_{\mathrm{w}}$ or $q_{\mathrm{in}}$. When $r = 1.0$, the barchan dune gives place to a symmetric longitudinal dune \citep{Parteli_et_al_2009}.

{\em{Model $\#2$}}: the primary wind has a larger shear stress (${\tau}_1 > {\tau}_2$) --- as in the conceptual model by \cite{Tsoar_1984} --- while $T_{{\mathrm{w}}1} = T_{{\mathrm{w}}2}$. In this case, the dune shapes do not differ much from the ones in Fig.~\ref{fig:duran}. The limb opposite to the secondary wind elongates when the relative values of bulk sand flux in the secondary and primary wind directions, respectively $Q_2$ and $Q_1$, are such that $r \equiv {Q_2}/{Q_1} > 25\%$. Indeed, we also considered the situation where the storm wind is the secondary one ($Q_2 > Q_1$), such as in the model by \cite{Bagnold_1941}. In this case, primary and secondary winds simply exchange their roles, whereas the same dune shapes such as in Fig.~\ref{fig:duran} are obtained. In summary, Bagnold's hypothesis that the elongating limb is the one exposed to a secondary storm wind is not supported by the simulation results.

In a general manner, the condition for elongation of the asymmetric limb due to a bimodal wind of obtuse divergence angle, as found from our calculations, reads:
\begin{equation}
r \equiv {\left[{Q_2 \cdot T_{{\mathrm{w}}2}}\right]}/{\left[{Q_1 \cdot T_{{\mathrm{w}}1}}\right]} > 1/4, \label{eq:R}
\end{equation}
where the indices 1 and 2 refer to the primary and to the secondary wind, respectively. Equation (\ref{eq:R}) allows us to estimate the onset for emergence of asymmetry in barchan dunes, where both the relative duration and shear stress values of the wind components are accounted for quantitatively. 

\begin{figure}[!t] 
\begin{center} 
\vspace{0.1cm}
\includegraphics[width=1.0\columnwidth]{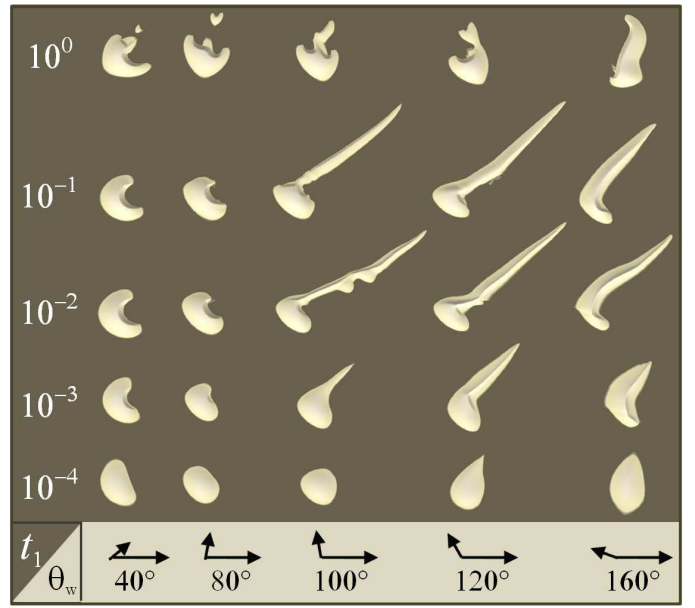} 
\caption{Dune shape as a function of the divergence angle (${\theta}_{\mathrm{w}}$) of the bimodal wind, and of the rescaled duration of the primary wind ($t_{1} \equiv T_{{\mathrm{w}}1}/T_{\mathrm{m}}$), for $r = 50\%$, where $r \equiv T_{{\mathrm{w}}2}/T_{{\mathrm{w}}1}$ is the ratio between the duration $T_{{\mathrm{w}}2}$ of the secondary wind and the duration of the primary wind ($T_{{\mathrm{w}}1}$). The directions of primary and secondary winds are indicated by the arrows (primary wind blows to the right). Simulation snapshots correspond to time $t \approx 3\,T_{\mathrm{m}}$, except those obtained with $t_1 = 10^{-4}$, which correspond to time $t \approx 0.2T_{\mathrm{m}}$.}
\label{fig:asymmetric_wind2} 
\end{center} 
\end{figure}

Next, we extend the calculations to different values of the divergence angle ${\theta}_{\mathrm{w}}$. In Fig.~\ref{fig:asymmetric_wind2}, the dune shape emerging from an asymmetric bimodal wind with $r = 50\%$ is shown as a function of ${\theta}_{\mathrm{w}}$ and $t_1 = T_{{\mathrm{w}}1}/T_{\mathrm{m}}$, which is the duration of the primary wind, $T_{{\mathrm{w}1}}$, rescaled by the migration or reconstitution time of the barchan, $T_{\mathrm{m}} \approx 0.02W^2/Q_1$ \citep{Duran_et_al_2010}. The dune turnover time is roughly the time needed for the dune to cover a distance of its width ($W$) \citep{Allen_1974,Lancaster_1988,Rubin_and_Ikeda_1990,Hersen_et_al_2004}. 

\begin{figure}[!t] 
\begin{center} 
\includegraphics[width=0.97\columnwidth]{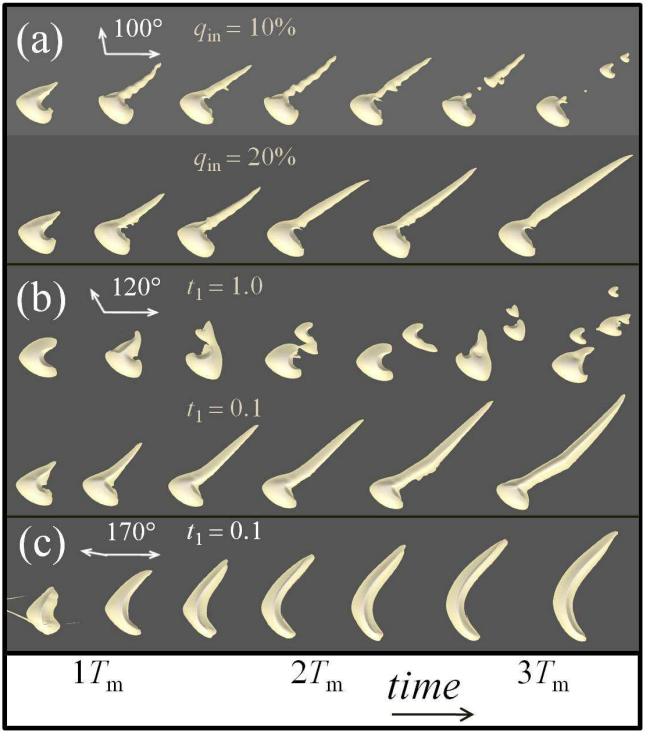}
\caption{Time evolution of the barchan shape under bimodal wind regimes with obtuse divergence angle and $r = 50\%$ (cf.~Fig.~\ref{fig:asymmetric_wind2}). Time is in units of the turnover time of the barchan ($T_{\mathrm{m}}$). (a) The divergence angle is ${\theta}_{\mathrm{w}} = 100^{\circ}$, the duration of the primary wind (to the right) relative to $T_{\mathrm{m}}$ is $t_1 = 0.1$, and the influx ($q_{\mathrm{in}}$) is $10\%$ (top) and $20\%$ (bottom); (b) ${\theta}_{\mathrm{w}} = 120^{\circ}$, $q_{\mathrm{in}} = 20\%$, $t_1 = 1.0$ (top) and $0.1$ (bottom); (c) ${\theta}_{\mathrm{w}} = 170^{\circ}$, $q_{\mathrm{in}} = 20\%$, $t_1 = 0.1$.}
\label{fig:asymmetric_wind3} 
\end{center} 
\end{figure}

As can be seen from Fig.~\ref{fig:asymmetric_wind2}, dunes orient transversely or longitudinally to the resultant transport trend, depending on whether ${\theta}_{\mathrm{w}}$ is acute or obtuse, respectively. Bimodal wind regimes with ${\theta}_{\mathrm{w}} < 90^{\circ}$ lead to rounded (or oblate) barchans, whereas larger values of ${\theta}_{\mathrm{w}}$ yield asymmetric dunes that elongate in the resultant transport direction \citep{Parteli_et_al_2009}. 
However, the elongating limb is not stable if ${\theta}_{\mathrm{w}}$ is within the range $90^{\circ} < {\theta}_{\mathrm{w}} \lesssim 112^{\circ}$. In this case, if the influx is sufficiently small, the extended limb separates from the barchan, giving rise to a ``mixed state'' \citep{Rubin_and_Hunter_1987,Rubin_and_Ikeda_1990,Parteli_et_al_2009}, as depicted in Fig.~\ref{fig:asymmetric_wind3}a. In fact, mixed states analogous to the ones found for asymmetric barchans have been also found in turntable experiments of aeolian ripples and subaqueous bedforms on a sand bed for the same range of ${\theta}_{\mathrm{w}}$ \citep{Rubin_and_Hunter_1987,Rubin_and_Ikeda_1990}. At very large divergence angles, the resulting morphology resembles an asymmetric dune of ``slim'' shape \citep{Long_and_Sharp_1964,Bourke_2010}, as shown in the diagram of Fig.~\ref{fig:asymmetric_wind2} and also illustrated with the calculation using ${\theta}_{\mathrm{w}} = 170^{\circ}$ in Fig.~\ref{fig:asymmetric_wind3}c. The dunes in Fig.~\ref{fig:asymmetric_wind3}c are ``reversing'' dunes, which are formed by winds coming from opposing directions. This dune form has been discussed by \cite{Tsoar_2001}, who also gave examples of reversing dunes occuring in South Africa \citep{Tsoar_2001}.

Asymmetric barchans such as the ones in Fig.~\ref{fig:duran} form when the relative duration of the primary wind ($t_1 \equiv T_{{\mathrm{w}}1}/T_{\mathrm{m}}$) is within the range $10^{-3} \lesssim t_1 \lesssim 10^{-1}$. When $t_1$ is smaller than about $0.1\%$, no limb elongation occurs; both limbs, as well as the slip face, disappear and a dome-like shape is obtained (cf.~Fig.~\ref{fig:asymmetric_wind2}). As $t_1$ approaches unity, any change due to the secondary wind is fully compensated by the primary wind during every cycle of the bimodal wind regime --- the resulting morphology is essentially a barchan that is periodically reformed by a unimodal wind of strength ${\tau}_1$, as can be seen from Fig.~\ref{fig:asymmetric_wind2}. The dune shapes shown in Fig.~\ref{fig:asymmetric_wind2} for $t_1 = 1.0$ are thus ``transitional'' shapes \citep{de_Hon_2006}. The oblique incidence of the secondary wind leads to destabilization of the dune surface and the emergence of smaller barchans, which detach from the larger dune and migrate on the bedrock thereby alternating between both directions of the bimodal wind. Figure \ref{fig:asymmetric_wind3}b shows the calculation of a barchan under asymmetric bimodal wind regime with ${\theta}_{\mathrm{w}} = 120^{\circ}$ for $t_1 = 1.0$ (top) and $t_1 = 0.1$ (bottom). On the basis of Fig.~\ref{fig:asymmetric_wind3}b, it is possible to understand the coexistence of barchans and other complex bedforms shaped by multimodal wind regimes. If the dune is small enough, it has a small turnover time ($\sim$ large $t_1$) and can readapt to the prevailing transport trend notwithstanding the complexity of the wind system \citep{Lancaster_1988}.

\section{\label{sec:topography}Topography}

Field evidence of barchan limb elongation caused by topography are unclear \citep{Finkel_1959,Long_and_Sharp_1964,Gay_1999,Bourke_2010}. We investigate two cases where topography is observed to trigger dune asymmetry under both constant wind direction and sediment supply. 

\begin{figure*}[!ht] 
\begin{center} 
\vspace{0.5cm}
\includegraphics[width=0.37\textwidth]{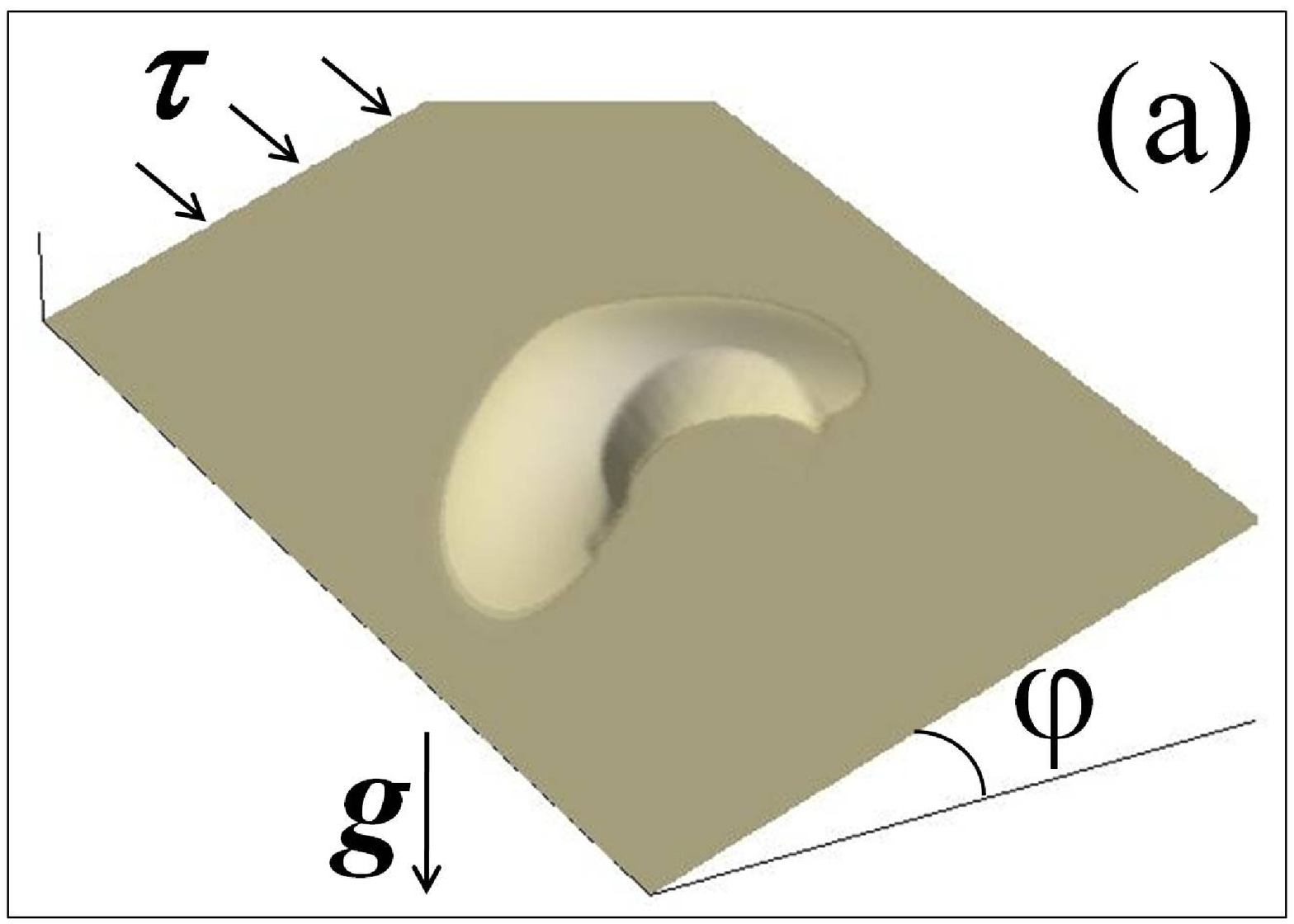}  
\includegraphics[width=0.48\textwidth]{fig5c.eps} 
\includegraphics[width=0.40\textwidth]{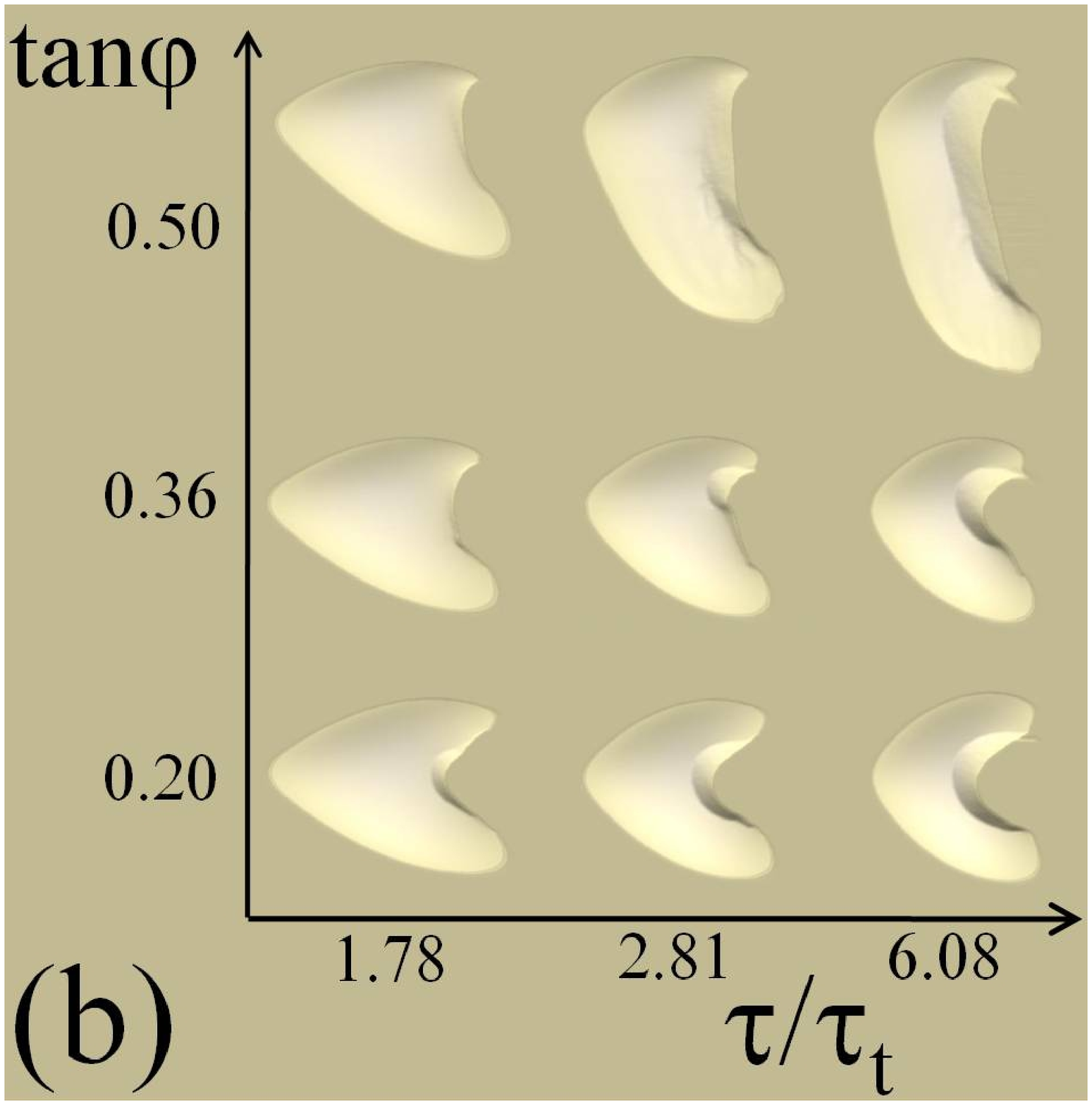}
\includegraphics[width=0.48\textwidth]{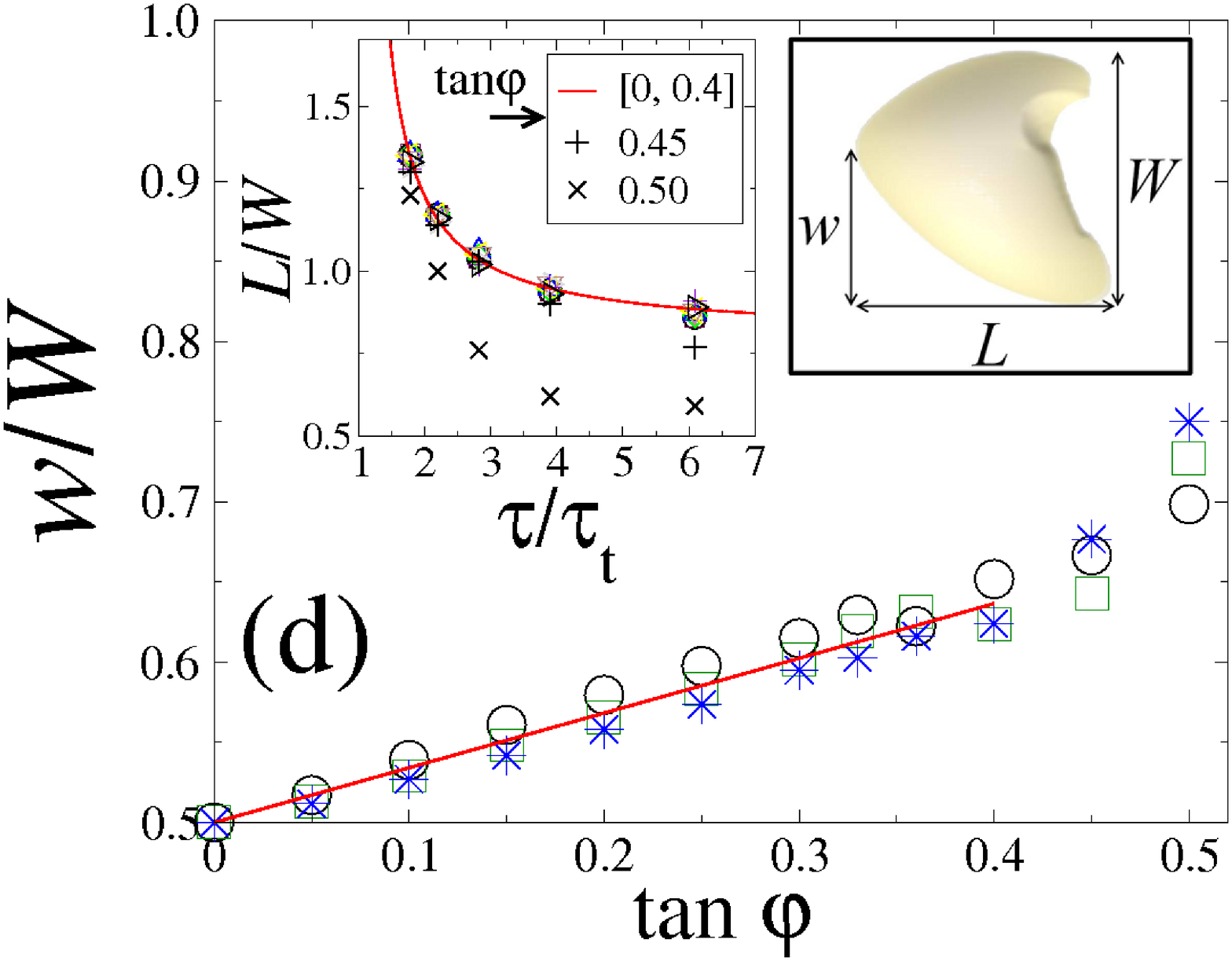} 
\caption{{\bf{(a)}} Barchan dune of cross-wind width $W \approx 90$ m migrating along a sloping terrain. The surface is tilted by an angle $\varphi$. Arrows indicate the wind direction (${\boldsymbol{{\tau}}}$ is the upwind shear stress) and gravity (${\boldsymbol{g}}$). {\bf{(b)}} Dune shape as a function of $\varphi$ and ${\tau}/{\tau}_{\mathrm{t}}$ (wind blows from left to right). Simulations were performed with $q_{\mathrm{in}}/q_{\mathrm{s}} = 0.2$. {\bf{(c)}} Main plot: Ratio between transverse and longitudinal migration velocities, respectively $v_{\mathrm{T}}$ and $v_{\mathrm{L}}$. The best fit to the data using the equation $v_{\mathrm{T}}/v_{\mathrm{L}} = k_q\,{\mbox{tan}}{\varphi}$ gives $k_q \approx 0.26$ (continuous line). Inset: $k_q$ as a function of $q_{\mathrm{in}}/q_{\mathrm{s}}$. {\bf{(d)}} Main plot: The degree of asymmetry is quantified in terms of the ratio $w/W$, calculated as a function of the surface inclination ($\varphi$) and for the same values of ${\tau}/{\tau}_{\mathrm{t}}$ shown in the main plot of (a). Inset: Length-to-width ratio ($L/W$) as a function of ${\tau}/{\tau}_{\mathrm{t}}$ calculated for different values of $\varphi$ within the range $0 < {\mbox{tan}}{\varphi} < 0.4$ and for two values of $\varphi$ above this range ($\varphi = 0.45$ and $\varphi = 0.50$). The best fit to the data within the range $0 < {\mbox{tan}}{\varphi} < 0.4$ using the equation $L/W \approx A{[{\tau}/{\tau}_{\mathrm{t}}-1]}^{-1} + B$ gives $A \approx 0.42$ and $B = 0.8$ (continuous line).}
\label{fig:topography1} 
\end{center} 
\end{figure*}

{\em{Sloping terrain}} --- We consider a barchan dune migrating on a sloping terrain that makes an angle $\varphi$ with the horizontal (Fig.~\ref{fig:topography1}a). The simulation starts with a symmetric barchan of height $H$ and width $W$, which was shaped on a flat surface under constant upwind shear stress ${\tau}$ and constant influx ($q_{\mathrm{in}}/q_{\mathrm{s}} = 0.2$). Then, the barchan is let to evolve on the inclined surface. Due to gravity, sand transport on the tilted surface has now a component perpendicular to wind trend.

Figure \ref{fig:topography1}b shows that the emerging barchan shape is asymmetric, yet none of its limbs elongate to form a longitudinal dune \citep{Parteli_et_al_2009,Reffet_et_al_2010}. The asymmetry arises from the combined effect of both downwind sand transport and gravity-driven mass flow in the direction orthogonal to the wind. Sand transport along the lower limb has a small net component downhill due to the tilting, and so this limb stretches downwhill. In contrast, the upper limb is not strechted because the sand transported laterally due to the tilting is trapped at the slip face.

A consequence of the surface inclination is the downhill extension of the barchan. The migration velocity of the barchan dune orthogonal to the wind trend increases with $\varphi$, as shown in the main plot of Fig.~\ref{fig:topography1}c. For moderate values of $\varphi$ within the range $0 \leq {\mbox{tan}}{\varphi} \leq 0.4$, the ratio between transverse and longitudinal migration velocities, $v_{\mathrm{T}}$ and $v_{\mathrm{L}}$, respectively, can be described by the equation:
\begin{equation}
v_{\mathrm{T}}/v_{\mathrm{L}} = k_q{\mbox{tan}}\varphi, \label{eq:vT_vL}
\end{equation}
where the coefficient $k_q \approx 0.26$ has a slight dependence on the upwind sand flux, $q_{\mathrm{in}}/q_{\mathrm{s}}$ (cf.~inset of Fig.~\ref{fig:topography1}c). The ratio $v_{\mathrm{T}}/v_{\mathrm{L}}$ for barchans migrating on a sloping terrain is thus a function of the slope $\varphi$, being independent of the wind velocity. 

The topography-induced barchan asymmetry can be quantified in terms of the relative width $w/W$, where $W$ is the total dune width, and the width $w$ is the distance --- measured in the direction orthogonal to wind trend --- between the border of the dune at the lowest elevation and the dune's windward foot (cf.~Fig.~\ref{fig:topography1}d). If there is no surface tilting ($\varphi = 0$), then $w/W = 0.5$ since the dune is symmetric along its central axis. As $\varphi$ increases within the range $0 \leq {\mbox{tan}}{\varphi} \leq 0.4$, $w$ increases roughly linearly with ${\mbox{tan}}\varphi$, as shown in Fig.~\ref{fig:topography1}d. Within this moderate range of $\varphi$, the ratio $w/W$ can be described by the equation,
\begin{equation}
w/W = k_w{\mbox{tan}}\varphi, \label{eq:wW}
\end{equation}
where the value $k_w \approx 0.33$ is obtained from the best fit to the simulation data (cf.~Fig.~\ref{fig:topography1}d). The aspect ratio of the dune ($L/W$), which is a function of the relative shear stress (${\tau}/{\tau}_{\mathrm{t}}$) \citep{Kroy_et_al_2002,Parteli_et_al_2007}, is essentially independent of $\varphi$. In other words, ${\tau}/{\tau}_{\mathrm{t}}$ and $\varphi$ can be approximately estimated from $L/W$ and $w/W$ of the asymmetric barchan shape, respectively, provided other asymmetry causes (i.e. bimodal wind, dune collisions or influx asymmetry) are not relevant at the dune field considered. 

\begin{figure}[!ht] 
\begin{center}
\includegraphics[width=1.00\columnwidth]{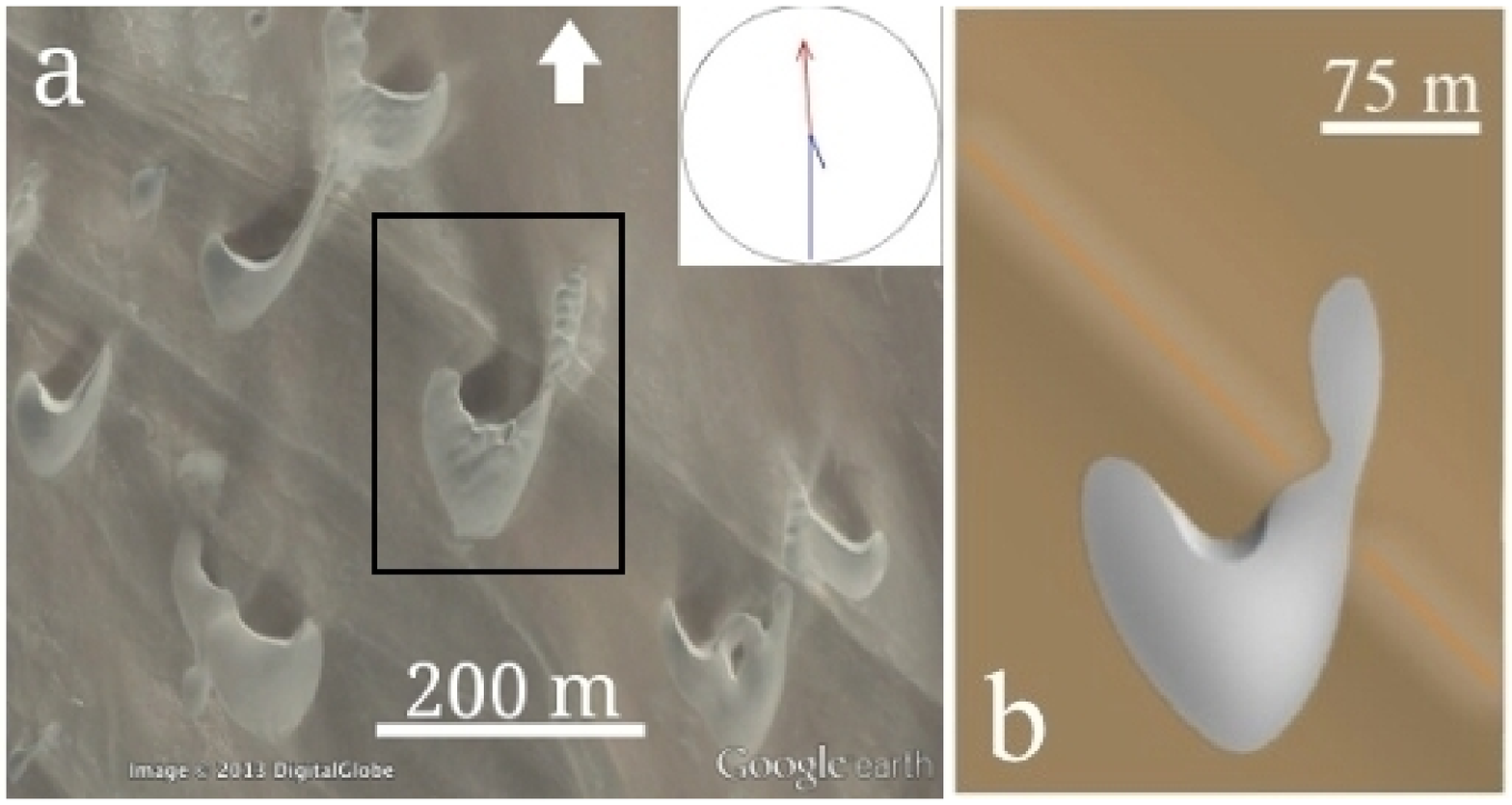}  
\includegraphics[width=1.00\columnwidth]{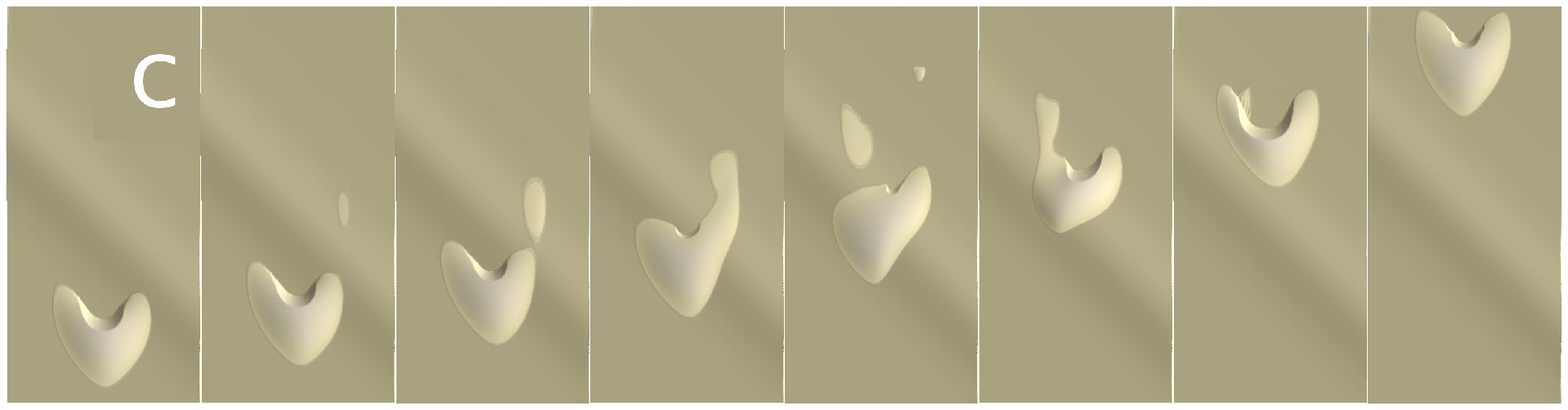}  
\caption{The potential influence of topography in the development of barchan asymmetry in Peru. (a) Barchan dunes migrating across a landscape where resistant layers in underlying bedrock protrude as low ridges. The dunes in the image are located near $14^{\circ}52^{\prime}26.43^{{\prime}\prime}$S, $75^{\circ}30^{\prime}58.31^{{\prime}{\prime}}$W. Illumination is from the top and the arrow indicates the north direction. The box indicates a barchan that extends its eastern limb across the topographic rise. The sand rose at the top-right corner is for San Juan de Marcona, located near $15^{\circ}21^{\prime}$S, $75^{\circ}09^{\prime}$W, about $68\,$km from the dunes in the image. Wind regime in the area is bimodal resulting in the extension of western limbs, except where influenced by topography. Image credit: Google Earth. (b) Asymmetric barchan dune of height $\sim 9$ m produced with the model using a topographic rise of Gaussian cross section which has along-wind width 200 m and height 6 m. (c) Simulation snapshots of the evolution of the barchan dune. Time increases from left to right and wind blows from the bottom. The simulation was performed with ${\tau}/{\tau}_{\mathrm{t}} \approx 3.9$ and $q_{\mathrm{in}}/q_{\mathrm{s}} = 0.2$.}
\label{fig:topographic_raise} 
\end{center} 
\end{figure}

{\em{Topographic rise}} --- Asymmetric barchans occasionally occur approaching and crossing a topographic break in slope \citep{Bourke_2010}. We investigate, using the dune model, the evolution of a barchan crossing a ridge that is placed obliquely to the direction of motion (Fig.~\ref{fig:topographic_raise}). The dune is subjected to an unimodal wind regime and an influx that is $20\%$ of the saturated flux. We consider that the longitudinal axis of the ridge forms a moderate angle with the direction orthogonal to the wind (about $45^{\circ}$). The ridge has Gaussian cross section, width $L_r$ and height $H_r$, while the barchan has a height $H \approx 9$ m. 

Indeed, the height-to-width ratio $H_r/L_r$ (or aspect ratio) of the ridge cannot be too large, as the ridge would act as a barrier for sand transport. On the contrary, a ridge of too small aspect ratio has negligible effect on the dune shape. An asymmetric dune shape with limb elongation occurs when the aspect ratio of the ridge is about $1/20$. In Figs.~\ref{fig:topographic_raise}b and \ref{fig:topographic_raise}c, the ridge has height $H_r = 7$ m and width $L_r \approx 150$ m. We have examined satellite images of the dune field of Fig.~\ref{fig:topographic_raise}a and noted that the cross section of the low ridges can indeed vary within the broad range between 50 and 150 m. Thus, both in the field and in the simulations, the ridge's cross section is of the same order of the cross-wind width of the barchan dune. As the barchan approaches the ridge, deposition occurs at the lee of the ridge, on the side corresponding to the closest limb. This limb then elongates once it arrives at the lee of the obstacle. The taller the ridge, the more accentuated the extension. As the other limb approaches sufficiently close to the ridge, a similar process occurs and thus that limb also elongates. 
\begin{figure*}[!ht] 
\begin{center}
\includegraphics[width=0.77\textwidth]{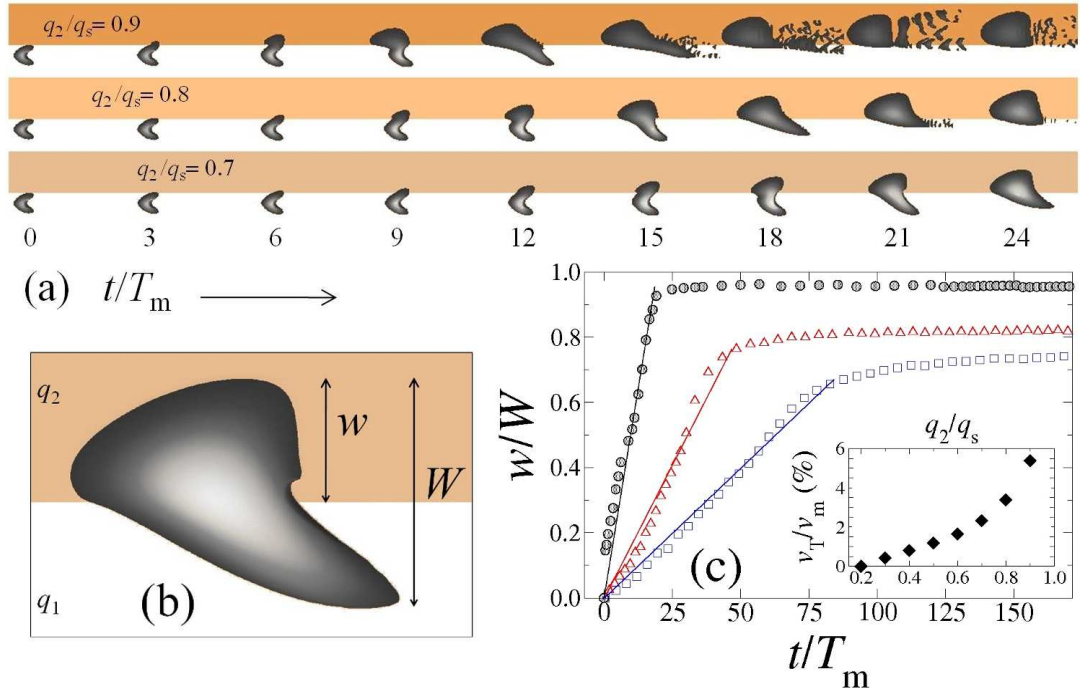} 
\caption{(a) Time evolution of a barchan of width 80 m subject to an asymmetric influx: The influx is $q_1 = 20\%$ of $q_{\mathrm{s}}$ in the white areas, in the dark areas the influx is $q_2$, where $q_2/q_{\mathrm{s}} = 90\%$, $80\%$ and $70\%$, respectively from top to bottom. The wind blows from the left with upwind shear stress ${\tau} = 3.24{\tau}_{\mathrm{t}}$. Snapshots at time interval ${\Delta}t = 3\,T_{\mathrm{m}}$ where $T_{\mathrm{m}}$ is the turnover time of the barchan at $t = 0$. (b) $w$ is the partial fraction of the barchan that is within the stream of larger influx ($q_2 = 0.7$, $t = 22\,T_{\mathrm{m}}$), whereas $W$ is the total width of the barchan. (c) Main plot: $w/W$ as a function of time for $q_2/q_{\mathrm{s}} = 0.4$ (squares), $0.6$ (triangles) and $0.9$ (circles). The straight lines are the best fits to the simulation data using a linear relation $w/W = kt/T_{\mathrm{m}}$, where $k = v_{\mathrm{T}}/v_{\mathrm{m}}$ and $v_{\mathrm{T}}$ denotes the rate with which $w/W$ changes with time in order to accomodate the asymmetric influx condition. The inset shows that $v_{\mathrm{T}}$ is less than $10\%$ of $v_{\mathrm{m}}$.} 
\label{fig:asymmetric_influx} 
\end{center} 
\end{figure*}

The deposition at the ridge's lee occurs due to a decrease in wind strength just downwind of the crest. When the sand released through the limbs crosses the ridge's crest, it cannot escape the lee since the transport rate there is reduced. As the dune advances and more limb sand is deposited, the wind strength at the lee suffices to extend the limb, yet accumulation ensues. Thus, from the sand released through this limb, a sand patch forms downwind of the ridge's crest (second frame from left to right), which develops into an elongated limb (cf.~third and fourth frames). Then, in the fifth frame, the left limb is crossing the ridge's crest which also leads to a sand patch in the ridge's wake. This process causes, then, elongation of the left limb as we can see in the subsequent frame.

It is interesting to note that in Fig.~\ref{fig:topographic_raise}a not all barchans elongate the same limb. For instance, some barchans which have the right (left) limb elongated are upwind (downwind) of the ridge's crest, which is in agreement with the prediction of Fig.~\ref{fig:topographic_raise}c as discussed above. However, this behavior can be explained only if we assume that the dunes are well isolated from each other (i.e. they are not participating at a collision). For instance, the dune shape in the most lower right of Fig.~\ref{fig:topographic_raise}a, which seems to result from a collision between dunes, has limbs of nearly equal size. Furthermore, the dune on the most lower left has an elongated arm opposite to that of the example barchan with the black box drawn around it. However, the elongated arm of the lower left dune too has a slip face --- possibly indicating that this dune too is being affected by a collision. Therefore, application of our results to the field should be made with care, since we did not account for the effect of dune collisions or influx asymmetry, which can influence the dune shape and will be addressed separately in later sections.

Future studies should focus on the asymmetric dune resulting from simulations using different angles between ridge orientation and wind direction, as well as different shapes of the ridge. Indeed, the results of our calculations should be valid for ridges which have a smooth cross section, where the slope of the surface does not exceed the critical angle for causing flow separation (about $20^{\circ}$) \citep{Sweet_and_Kocurek_1990}. If the ridge's cross section has a sharp edge, then flow separation occurs and the flow is diverted parallel to the ridge along the lee side \citep{Tsoar_2001}. In this case, sand can be transported in the direction parallel to the ridge wherever the wind speed within the separation zone exceeds the transport threshold. Moreover, the size of the zone of recirculating flow increases with the crest-to-brink distance of transverse rigdes \citep{Schatz_and_Herrmann_2006}, and thus the effect of secondary flow on the lateral transport of sand at the lee of the ridge should also depend on the shape of the ridge at its windward side. However, due to the assumption of our model that net flow ceases within the separation zone, the sand deposited at the lee of a ridge with sharp edge cannot be transported at all. Thus, three-dimensional flow patterns at the lee of dunes or obstacles should be accounted for in the modeling of dune asymmetry when the topographic rise has a sharp slope \citep{Tsoar_2001}. 


\section{\label{sec:asymmetric_sand_supply}Asymmetric sand supply}

The role of asymmetric sediment supply for barchan asymmetry has remained uncertain in the few investigations undertaken \citep{Rim_1958,Lancaster_1982,Bourke_2010}. It is admittedly a difficult task to conduct systematic field studies on the barchan shape as a function of the degree of spatial asymmetry in the upwind sediment budget. The sand flux onto the windward side of a barchan is largely dependent upon the spatial distribution of the upwind dunes or other sand sources, e.g. at a sand sea margin \citep{Lancaster_1982}. Spatial inhomogeneities in the local physical properties of the interdune terrain may also play an important role for the incoming flux \citep{Fryberger_et_al_1984}. Furthermore, the shape of a barchan migrating in a field is inevitably affected by wind trend variations in time \citep{Elbelrhiti_et_al_2005}, as well as by the interaction with neighbouring dunes through merging, breeding and lateral linking.

With the help of simulations, the barchan shape resulting from asymmetric upwind flux alone can be studied without regard to other physical factors. We consider a symmetric barchan initially migrating within an area where the influx $q_1$ is approximately a constant fraction ($\approx 20\%$) of the saturated flux ($q_{\mathrm{s}}$) --- the value of $q_1/q_{\mathrm{s}} \approx 0.2$ is, in fact, within average values measured at interdune areas in typical barchan dune fields \citep{Fryberger_et_al_1984}. The barchan has height $H \approx 6.0$ m and width $80$ m, and has been shaped by a constant shear stress $\tau \approx 3.24{\tau}_{\mathrm{t}}$. 

At $t=0$, the upwind influx becomes asymmetric. The influx on the side of the barchan's left limb has a larger value, $q_2 > q_1$ (dark areas in Fig.~\ref{fig:asymmetric_influx}a). This model for asymmetric influx constitutes a simple model for a scenario typical for real dune fields, where in some areas the presence of a dome, a sand sheet or a larger dune upwind may significantly increase the local interdune flux. Figure \ref{fig:asymmetric_influx}a shows different snapshots of the evolution of the barchan dune subject to the asymmetric influx. Because more sand deposits in the areas of larger influx ($q_2$), the dune adapts to the asymmetric influx by increasing its relative volume within the areas under influx $q_2$. Different (transient) asymmetric dune shapes appear as the dune reforms its shape in order to accomodate the larger influx $q_2$, as can be seen in Fig.~\ref{fig:asymmetric_influx}a. The shorter limb is the one exposed to the larger influx value, because a large upwind flux prevents erosion at the windward foot and downwind motion of the dune. Thus, the barchan limb on the side of lower influx appears advanced in relation to its --- ``fat'' \citep{Long_and_Sharp_1964,Parteli_et_al_2007} --- counterpart. 

The fraction of the dune width ($w/W$, cf.~Fig.~\ref{fig:asymmetric_influx}b) that is within the area of larger influx ($q_2$) increases in time, as can be seen in Fig.~\ref{fig:asymmetric_influx}c. In this figure, $v_{\mathrm{T}}$ denotes the rate with which $w/W$ grows in time. Firstly, $w/W$ increases roughly as a linear function of time which means that in this initial stage $v_{\mathrm{T}}$ is approximately constant for given values of $q_1$ and $q_2$ relative to $q_{\mathrm{s}}$. The inset of Fig.~\ref{fig:asymmetric_influx}c shows that $v_{\mathrm{T}}$ is indeed much smaller than the longitudinal migration velocity of the barchan ($v_{\mathrm{m}}$) even for large $q_2/q_{\mathrm{s}}$. Indeed, the main plot of Fig.~\ref{fig:asymmetric_influx}c suggests that, after a sufficiently long time, $w/W$ asymptotically approaches a maximal value (and thus $v_{\mathrm{T}} \rightarrow 0$). Since the increase of $w/W$ with time becomes extremely slow after long time, and since the dune is increasing rapidly in volume as $q_2$ approaches $q_{\mathrm{s}}$, it is difficult to reliably estimate the steady-state value of $w/W$ through numerical simulations. However, we can approximately estimate the steady-state value of $w/W$ (which we call $w_{\infty}/W$) by means of a simple calculation, as we will explain in the next paragraph.

\begin{figure}[!ht] 
\begin{center} 
\includegraphics[width=0.9\columnwidth]{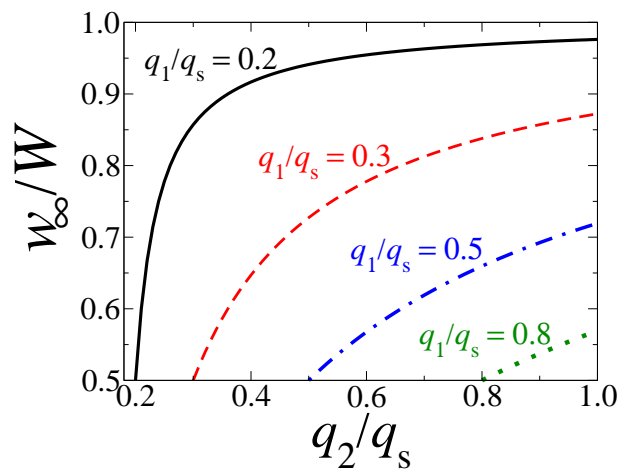}
\caption{Steady-state value $w_{\infty}/W$ of the fraction $w/W$ of the dune cross-wind width within the region of larger influx ($q_2$), as a function of $q_2/q_{\mathrm{s}}$ for different values of $q_1/q_{\mathrm{s}}$.} 
\label{fig:asymmetric_influx2} 
\end{center} 
\end{figure}

A barchan under constant influx $q_{\mathrm{in}}$ is an unstable object. While the dune is gaining sand due to $q_{\mathrm{in}}$, it is also losing mass through the limbs. The mass balance determines whether the dune grows or shrinks with time. As shown previously \citep{Duran_et_al_2010}, there is a critical value of $q_{\mathrm{in}}$, denoted by $q_{\mathrm{c}}$, above which the dune volume increases in time and below which the dune shrinks. This critical influx is approximately equal to $18\%$ of the saturated flux, $q_{\mathrm{s}}$ \citep{Duran_et_al_2010}. The rate ${{\mbox{d}}W}/{{\mbox{d}}t}$ at which the cross-wind width $W$ of the barchan changes in time approximately scales with ${\left[{q_{\mathrm{in}}-q_{\mathrm{c}}}\right]}/W$ \citep{Duran_et_al_2010}. Using this relation, we can approximately estimate the growth rate of $w$ and of $W-w$, which denote the fractions of the dune width under influx values $q_2$ and $q_1$, respectively (cf.~Fig.~\ref{fig:asymmetric_influx}b). The growth rates ${{\mbox{d}}w}/{{\mbox{d}}t}$ and ${{\mbox{d}}(W-w)}/{{\mbox{d}}t}$ should be proportional to ${\left[{q_2-q_{\mathrm{c}}}\right]}/w$ and ${\left[{q_1-q_{\mathrm{c}}}\right]}/(W-w)$, respectively. The steady-state value of $w$, i.e. $w_{\infty}$ is achieved when both rates equal, which gives, 
\begin{equation}
\frac{w_{\infty}}{W} = \frac{q_2 - q_c}{q_1+q_2-2 \cdot q_c},
\end{equation}
with $q_{\mathrm{c}} \approx 0.18\,q_{\mathrm{s}}$ as mentioned above. Figure \ref{fig:asymmetric_influx2} shows how $w_{\infty}/W$ depends on $q_2/q_{\mathrm{s}}$ for different values of $q_1/q_{\mathrm{s}}$, with $q_1,q_2 > q_{\mathrm{c}}$. For values of $q_1$ close to $q_{\mathrm{c}}$, $w_{\infty}/W$ approaches unity as $q_2$ increases, which means that the dune enters almost entirely into the region of larger influx and is nearly symmetric again. This situation is indeed observed in the simulation (see e.g.~Fig.~\ref{fig:asymmetric_influx}a for $q_2/q_{\mathrm{s}} = 0.9$). However, as the influx becomes very large the barchan shape gives place to a flat dome-like dune without slip-face, which may serve as source of sand for smaller dunes \citep{Duran_et_al_2010,Luna_et_al_2011,Luna_et_al_2012}. We note that an increase of $q_1$ implies a decrease in the influx asymmetry, since $q_2$ is within the range $q_1<q_2<q_{\mathrm{s}}$. Thus, as $q_1$ increases, the steady-state shape of the dune becomes less symmetric (though the barchan shape gives place to a dome as discussed above) and $w_{\infty}/W \rightarrow 0.5$. Evidently, for $q_1 = q_2$ there is no asymmetry in the influx and the dune shape is thus symmetric ($w_{\infty}$ is equal to $W/2$). 

\section{\label{sec:dune_collision}Dune collisions}

\begin{figure}[!ht] 
\begin{center}
\includegraphics[width=0.6\columnwidth]{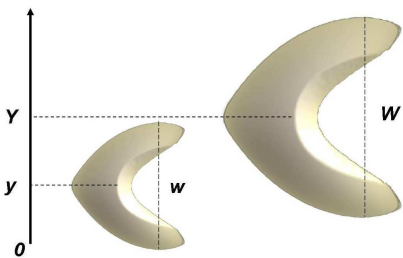} 
\caption{Binary collision between barchans of different sizes. A small barchan (width $w$) approaches a larger one (width $W$) from behind with lateral offset. The small and the large barchans are centered at $y$ and $Y$, respectively.} 
\label{fig:dune_collision1} 
\end{center} 
\end{figure}

Dune collisions are well-known factor for the emergence of complex, asymmetric dune morphologies \citep{Tsoar_2001,Bourke_2010,Ewing_and_Kocurek_2010}. The dynamics of collisions between two barchans have been systematically investigated in previous studies using the present model \citep{Schwaemmle_and_Herrmann_2003,Duran_et_al_2005,Duran_et_al_2009}. It was shown that, if the symmetry axes of both dunes are aligned, then the collision dynamics is determined by the volume ratio of the dunes, whereas the resulting morphology is symmetric \citep{Duran_et_al_2005}. However, if the collision occurs with a lateral offset (cf.~Fig.~\ref{fig:dune_collision1}), asymmetric dune shapes can arise \citep{Close_Arceduc_1969,Hersen_and_Douady_2005,Duran_et_al_2009}. 

\begin{figure}[!ht] 
\begin{center} 
\includegraphics[width=1.0\columnwidth]{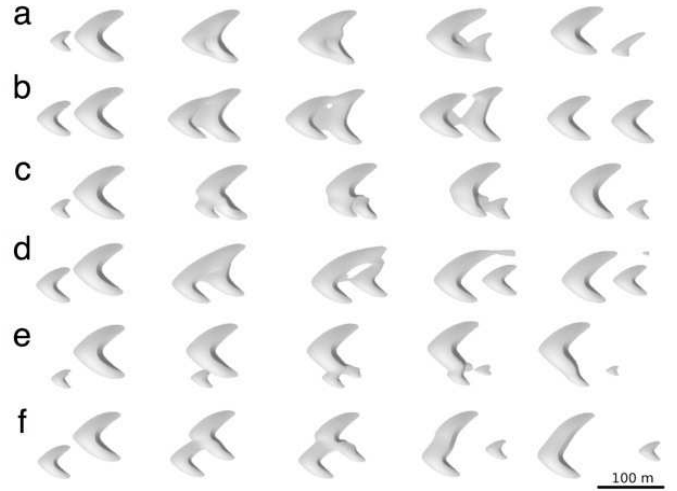}
\caption{Snapshots of simulations of binary collisions for different values of $\Delta$ and volume ratio $r \propto (w/W)^3$. Calculations with $r = 0.06$ are shown for (a) $\Delta = 0.2$, (c) $\Delta = 0.6$ and (e) $\Delta = 0.9$; calculations with $r = 0.3$ are shown for (b) $\Delta = 0.2$, (d) $\Delta = 0.6$ and (f) $\Delta = 0.9$.}
\label{fig:dune_collision2} 
\end{center} 
\end{figure}

Figure \ref{fig:dune_collision2} shows snapshots of calculations using different values of volume ratio and lateral offset --- defined as ${\Delta} = |Y-y|/(W/2)$, where $Y$ ($y$) is the crest's position of the large (small) dune in the direction orthogonal to the wind, and $W$ is the width of the large dune (Fig.~\ref{fig:dune_collision1}). The asymmetric dune shapes produced due to collisions are different from the ones in Section \ref{sec:asymmetric_sand_supply}, since the collision not only implies an asymmetric influx but also leads to a (dynamic) modification in the shear stress field during the interaction between the dunes \citep{Duran_et_al_2005}. The hybrid, asymmetric dunes depicted in Fig.~\ref{fig:dune_collision2} are transient shapes. A sufficiently long time after the collision, dunes are well separated and adapt to constant influx and wind direction again. This time-scale which the dune needs to adapt its shape after the collision is governed by the dune reconstitution time, $T_{\mathrm{m}}$, which scales with the barchan's cross-wind width $W$ divided by its migration velocity $v_{\mathrm{m}}$ \citep{Hersen_et_al_2004,Duran_et_al_2010}. Within this time-scale, the dune migrates a downwind distance of the order of its own width $W$.

The simulations corroborate previous observations that dune asymmetry could result from limb merging as two barchans link laterally, thereby causing downwind extension of the coalesced limbs \citep{Bourke_2010}. Such a situation is found for sufficiently large offset ($\Delta > 0.9$), as can be seen in Figs.~\ref{fig:dune_collision2}e and ~\ref{fig:dune_collision2}f. It is interesting to notice that the result of the lateral coalescence is the release of a small dune from the merged limbs, whereas the colliding barchans merge then into a single asymmetric dune. The release of the small dune from the merged limbs is signature of the so-called ``solitary wave-like behaviour'' of sand dunes discussed in previous studies of binary collisions between barchans \citep{Schwaemmle_and_Herrmann_2003,Duran_et_al_2005}. As the smaller, faster dune approaches the larger one from the upwind, sand from the downwind dune is trapped in the wake of its companion. Thus, the smaller dune upwind gains sand from the downwind dune thereby increasing in volume and becoming slower. Due to erosion downwind of the smaller dune's wake region (i.e. after the ``separation bubble'' of the small dune), a small dune is released from the larger barchan. In effect what happens is that the colliding dunes merge to form a single (asymmetric) dune while a small barchan is ejected from the surface of the downwind dune.

Moreover, we see in Figs.~\ref{fig:dune_collision2}e,f that the volume ratio of dunes colliding with large offset can yield distinct asymmetric dune shapes. If the upwind dune is small, it merges laterally to the lowest portions of the downwind barchan's limb which is involved in the collision (cf.~Fig.~\ref{fig:dune_collision2}e). This limb appears then deformed (with a somewhat ``sinuous'' shape) and its cross-wind width also increases. However, no significant limb extension in the wind direction occurs, as we can see in the last frame of Fig.~\ref{fig:dune_collision2}e. A different situation occurs if the size of the upwind dune participating at the collision is large, as in the simulation of Fig.~\ref{fig:dune_collision2}f. As we can see in the last frame of this figure, the upper limb of the larger barchan left behind (i.e. the limb {\em{opposite}} to the side where the binary collision occurred) appears {\em{advanced}} downwind. The reason for this behaviour is that the lower limb is much larger and slower than the upper one, thus appearing ``retarded'' in relation to its companion.

\begin{figure*}[!ht] 
\begin{center}
\includegraphics[width=0.9\textwidth]{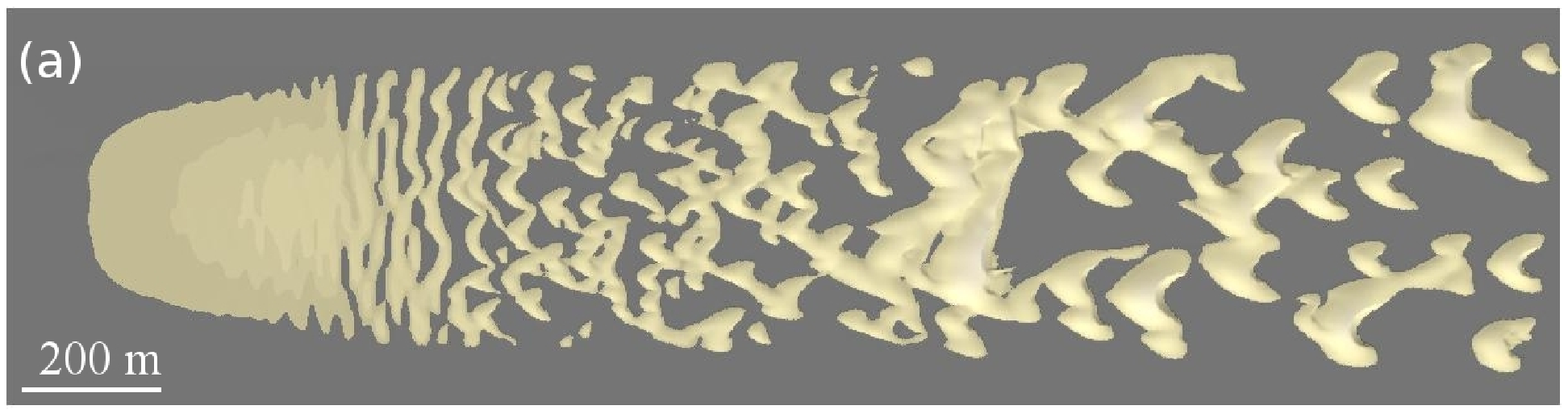}
\includegraphics[width=0.9\textwidth]{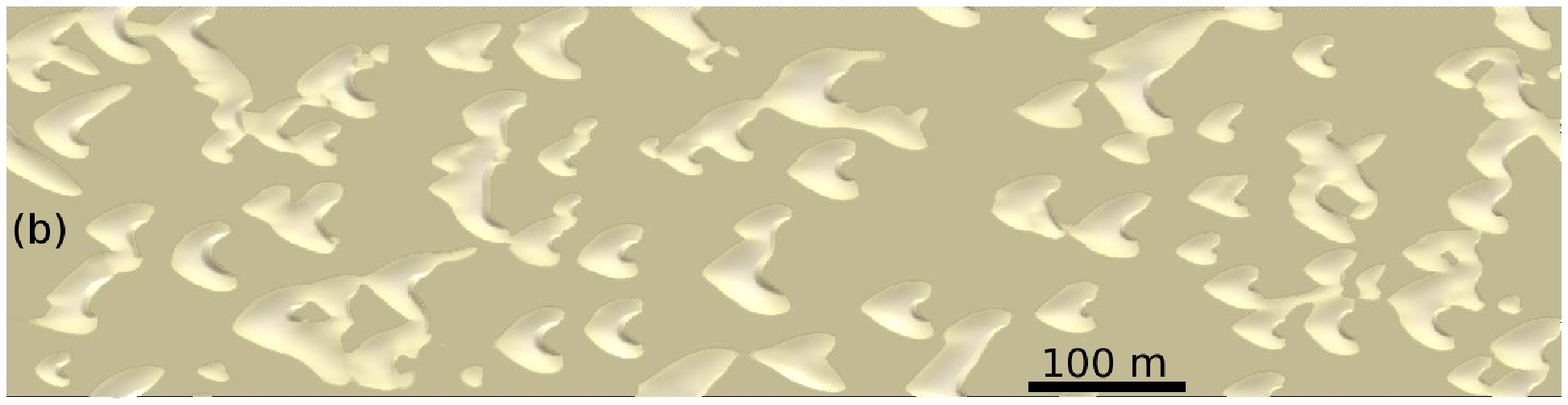}
\includegraphics[width=0.9\textwidth]{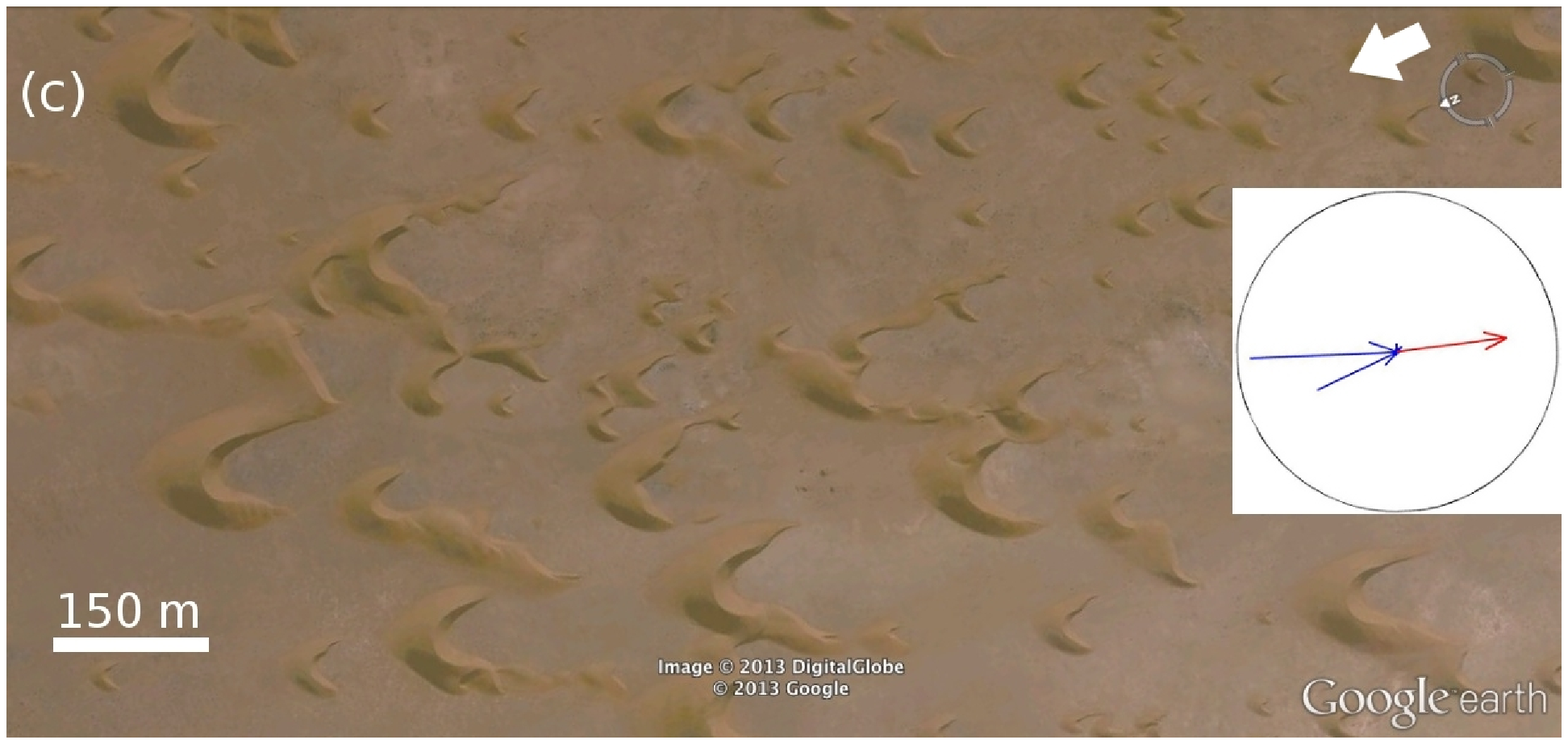} 
\caption{Asymmetric barchans emerge during the evolution of a barchan dune field. (a) Simulation of the genesis of a dune field. A small, flat sand hill subjected to saturated flux becomes a source of sand for a barchan dune field \citep{Duran_et_al_2010,Luna_et_al_2012}. Collision between multiple dunes lead to complex asymmetric dune morphologies which are not obtained in binary collisions (cf.~Fig.~\ref{fig:dune_collision2}). (b) Complex asymmetric barchan shapes also occur in the simulation of a barchan corridor using periodic boundary conditions both in the direction perpendicular and parallel to the wind. Wind blows from left to right in both (a) and (b). Simulations were performed with ${\tau}/{\tau}_{\mathrm{t}} \approx 3.9$. (c) Barchan dunes in Morocco, located at $26^{\circ}47^{\prime}02.81^{{\prime}{\prime}}$N, $13^{\circ}22^{\prime}42.16^{{\prime}{\prime}}$W (image credit: Google Earth). The north direction is indicated by the arrow at the top-right corner. The sand rose in the inset at right is for El Aaiun, which is located near $27^{\circ}09$N, $13^{\circ}12$W, about $42\,$km from the dunes in the image.}
\label{fig:barchan_field} 
\end{center} 
\end{figure*}

Asymmetric dune patterns very different from the ones in Fig.~\ref{fig:dune_collision2} can result from collisions between multiple dunes. Figure \ref{fig:barchan_field}a shows one snapshot of the simulation of a barchan dune field emerging from a flat sand surface subjected to a saturated influx. The simulations of the genesis of barchan fields were discussed in detail in previous works \citep{Duran_et_al_2010,Luna_et_al_2011,Luna_et_al_2012}. As shown in these works, the small transverse bedforms emerging from the instabilities which develop on the flat hill upwind give place to small barchans after reaching the bedrock --- due to the cross-wind instability inherent to transverse dunes \citep{Reffet_et_al_2010,Parteli_et_al_2011,Melo_et_al_2012,Niiya_et_al_2012}. The barchan dunes, which are subjected to a strong influx, grow in size during their migration, thereby experiencing multiple collisions with their counterparts and producing complex, asymmetric dune shapes. Different patterns produced in the simulation include limb extension of the upwind barchan, as we can see in Fig.~\ref{fig:barchan_field}a. While the simulation of Fig.~\ref{fig:barchan_field}a denotes a model for an immature dune field (which is at the early stages of its development), we have also performed simulations using periodic boundary conditions in the wind direction such as to model the steady-state dynamics of a dune corridor \citep{Duran_et_al_2010}. Such simulations also lead to asymmetric barchan shapes which cannot be obtained with binary collisions (cf.~Fig.~\ref{fig:barchan_field}b).

The complex asymmetric shapes shown in Figs.~\ref{fig:barchan_field}a and ~\ref{fig:barchan_field}b result from collisions between multiple barchans which have different sizes and offset values, and are further subjected to an asymmetric influx depending on the spatial distribution of dunes upwind. The systematic study of collisions between multiple dunes in a field is out of the scope of the present study. However, on the basis of our calculations, we can conclude that binary collisions can produce a rather limited range of asymmetric dune patterns. The only previously reported asymmetric pattern produced by a binary collision is the limb elongation due to dunes merging laterally (Fig.~\ref{fig:dune_collision2}e,f). Other asymmetric patterns induced by collisions can result if the collisions involve multiple dunes during the evolution of a barchan dune field. As can be seen in Figs.~\ref{fig:barchan_field}a and ~\ref{fig:barchan_field}b, collisions between dunes lead to the formation of groups of barchans, in which upwind barchans releasing sand through their extended limbs act as source of sediment to the dunes in the front, as previously noted by \cite{Tsoar_2001}. Indeed, the complex patterns emerging from collisions between barchans are transient dune shapes, as shown in Fig.~\ref{fig:dune_collision2} and in previous works \citep{Schwaemmle_and_Herrmann_2003,Duran_et_al_2005,Duran_et_al_2010}. Further modeling work is required in order to deepen our understanding of the collision dynamics of multiple dunes and of the resulting dune patterns.

\section{\label{sec:discussion}Discussion}

The results of our simulations can help researchers in the future to discern among the most relevant factors competing for diverse types of barchan asymmetry.

For instance, our model predicts that only asymmetric bimodal wind regimes can lead to seif dune formation due to an elongating barchan limb (cf.~Section \ref{sec:bimodal_winds} and the condition for seif dune elongation in Eq.~({\ref{eq:R}})). Barchans subjected to an asymmetric influx, crossing a topographic rise or undergoing a collision with other dunes can experience different types of limb extension, however a seif dune cannot appear from an asymmetric barchan due to these factors alone. Furthermore, our simulations show that topographic breaks in slope and collisions can trigger extension of one or the other limb depending on the location of the dune relative to the topographic obstacle or to the surrounding dunes (cf.~Figs.~\ref{fig:topographic_raise}, \ref{fig:dune_collision2} and \ref{fig:barchan_field}). In contrast, dunes subjected to the same asymmetric bimodal wind regime should elongate the same limb, provided other asymmetry causes are not influencing the dune morphology.

Application of the results of this paper to the field should be made with care because typically asymmetric barchans result from the concurrent action of more than one asymmetry causes. In the present paper, we list three examples of barchan dunes displaying different asymmetric shapes, the origin of which could be attributed to one specific asymmetry cause, based on our simulations. These examples are dicussed separately in Sections \ref{sec:example_1}$-$\ref{sec:example_3}, while in Section \ref{sec:planetary_discussion} we discuss on how our findings could help the research of extraterrestrial dunes.

\subsection{\label{sec:example_1}First example: barchan asymmetry induced by an assymmetric bimodal wind regime}

\begin{figure}[!t] 
\begin{center} 
\vspace{0.1cm}
\includegraphics[width=0.9\columnwidth]{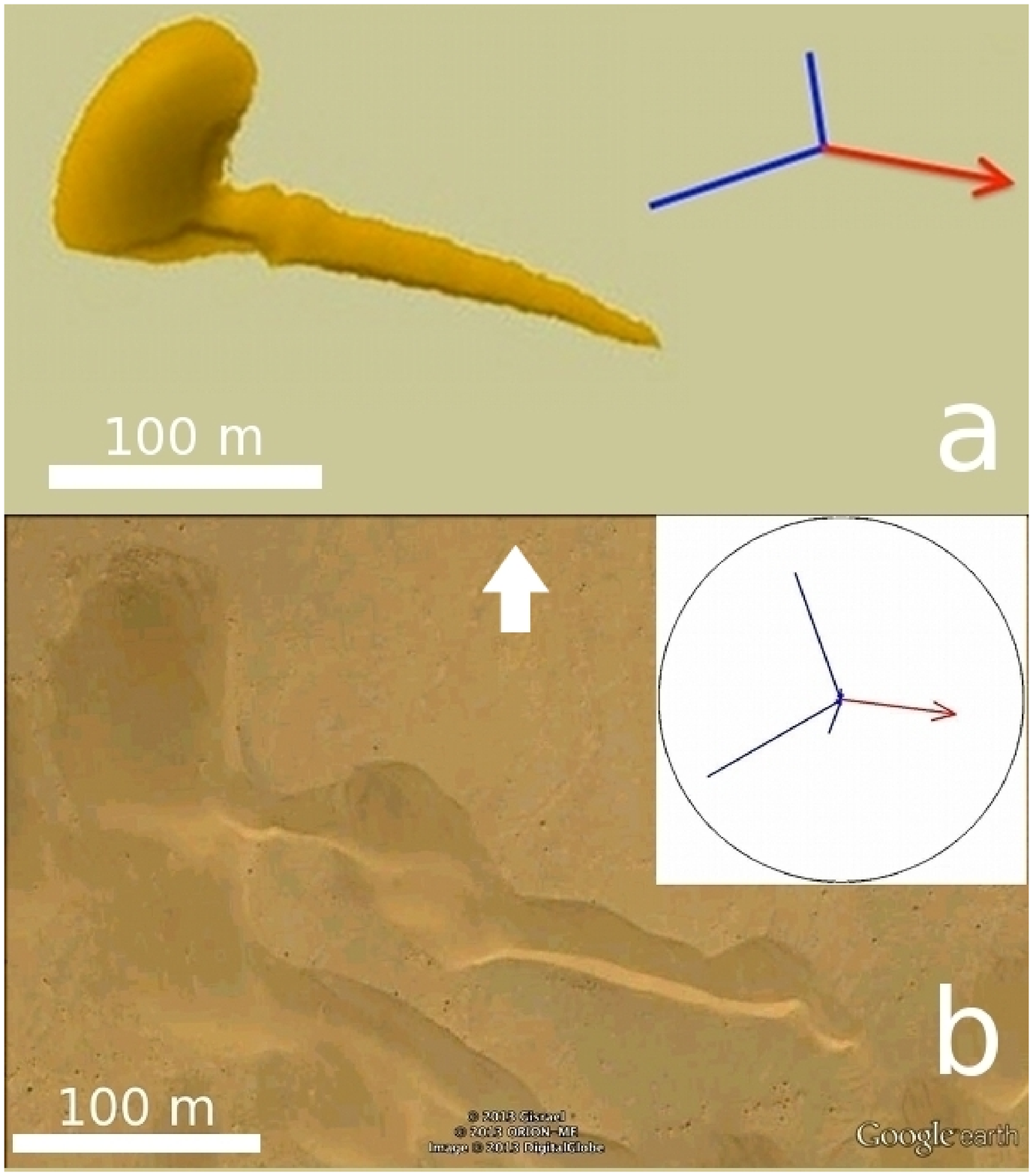} 
\includegraphics[width=0.9\columnwidth]{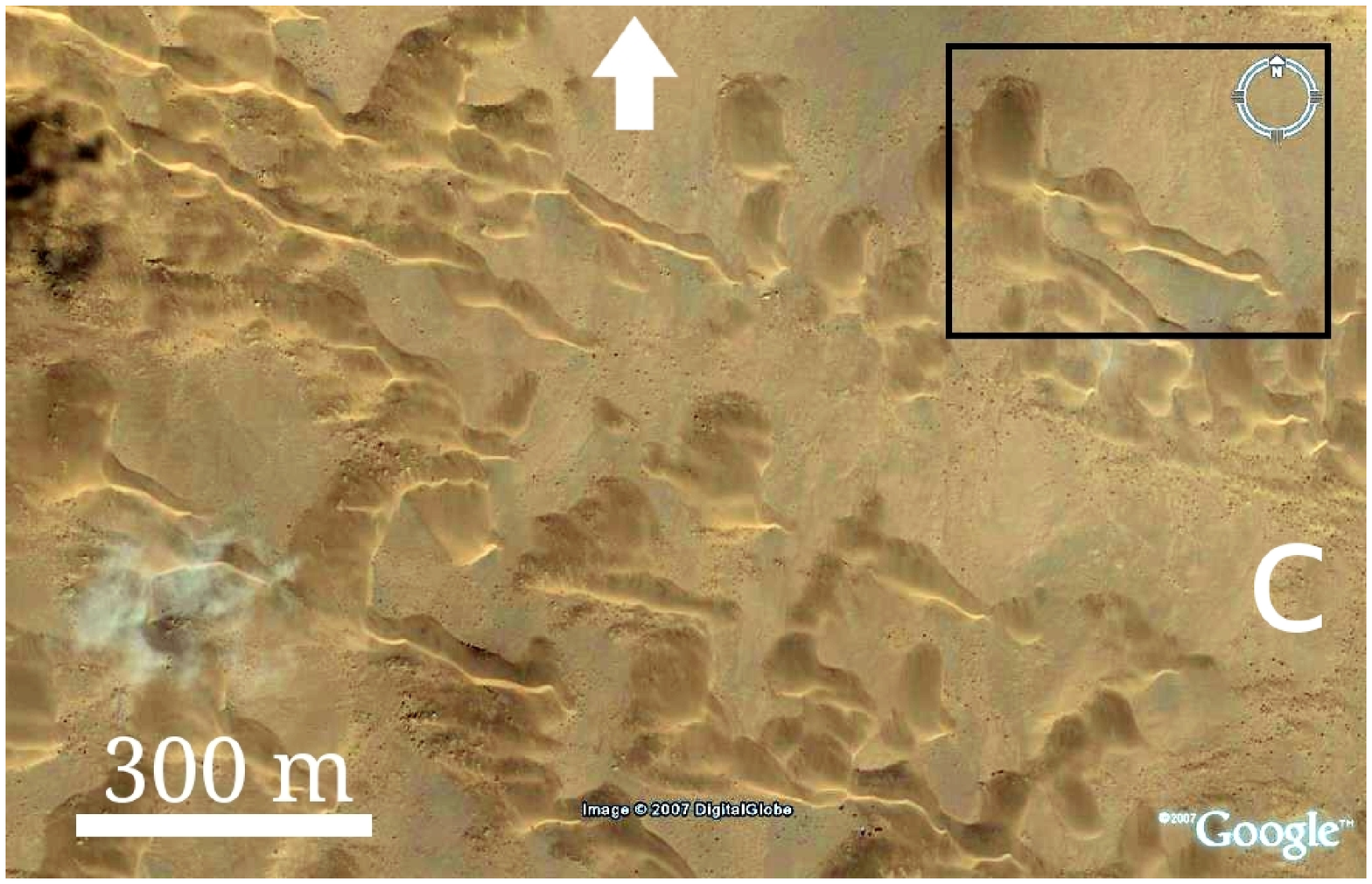} 
\caption{Barchan asymmetry due to an asymmetric wind regime. In {\bf{(a)}} we see a dune obtained in simulation using a bimodal wind regime with $r = 50\%$ and $t_1 = 0.1$ (see Fig.~\ref{fig:asymmetric_wind3}a). The influx is $q_{\mathrm{in}} = 20\%$ of the saturated flux. We compare this dune with the asymmetric barchan in {\bf{(b)}}, which is at Bir Lahfan in Sinai, near $30^{\circ}51^{\prime}04.03^{{\prime}{\prime}}$N, $33^{\circ}52^{\prime}24.19^{{\prime}{\prime}}$E. The sand rose shown in the inset is for an area between Bir Lahfan and Jabel Libni, Sinai. {\bf{(c)}} Image of the dune field at Bir Lahfan showing that the dune in (b), enclosed by the box, is surrounded by other asymmetric barchans which are also giving rise to seif dunes. Both in (b) and (c), north is at top as indicated by the arrows (images credit: Google Earth).}
\label{fig:comparison_bimodal_wind} 
\end{center} 
\end{figure}

The dune shown in Fig.~\ref{fig:comparison_bimodal_wind}b is in Bir Lahfan and provides an example of seif dune formation from a barchan dune due to obtuse bimodal wind regime \citep{Tsoar_1978,Tsoar_1983,Tsoar_1984,Tsoar_2001}. The sand rose in the inset shows two main wind directions with a divergence angle about $100^{\circ}$. We see that the dune displays only one limb, which aligns nearly longitudinally to the resultant transport trend. The growth and migration of the longitudinal seif dunes in Bir Lahfan was discussed in detail in previous publications (see e.g.~\cite{Tsoar_1978} for a review).

The dune in Fig.~\ref{fig:comparison_bimodal_wind}a was obtained from a simulation using an asymmetric  bimodal wind regime with $r = 50\%$. It is the third dune from left to right in Fig.~\ref{fig:asymmetric_wind3}a bottom ($q_{\mathrm{in}} = 20\%$ of the saturated flux). This dune indeed displays some characteristics that are found in the real Bir Lahfan dune, namely it has only one limb, which is elongating into a seif dune. 

The wind regime that generated the dune of Fig.~\ref{fig:comparison_bimodal_wind}a differs from the one in the sand rose of Fig.~\ref{fig:comparison_bimodal_wind}b as in this sand rose both main wind directions have nearly equal transport rates. Indeed, it was shown in experiments \citep{Reffet_et_al_2010} and numerical simulations \citep{Parteli_et_al_2009}, that symmetric bimodal flow regimes (with $r = 100\%$) lead to {\em{symmetric}} longitudinal dunes rather than asymmetric barchans. However, we note that the sand rose in Fig.~\ref{fig:comparison_bimodal_wind}b also displays secondary transport directions, which are not taken into account in the simulations. The role of these smaller transport directions for the dune shape is still uncertain. Simulations using multidirectional wind regimes, which are out of the scope of the present work, should be performed in the future in order to clarify the relevance of secondary transport directions for the shape and elongation of asymmetric barchans.

Possibly, the presence of more than two wind directions as well as influx asymmetry due to the presence of other dunes in the field (see Fig.~\ref{fig:comparison_bimodal_wind}c) could explain the meandering of the barchan limb visible in Fig.~\ref{fig:comparison_bimodal_wind}b. This meandering is namely absent from the asymmetric barchan of Fig.~\ref{fig:comparison_bimodal_wind}a and also from symmetric longitudinal dunes on top of bedrock generated in experiments \citep{Reffet_et_al_2010} and numerical simulations \citep{Parteli_et_al_2009}.

\subsection{\label{sec:example_2}Second example: barchan asymmetry induced by topography}

The second example is provided in Fig.~\ref{fig:topographic_raise}a, which shows asymmetric barchans in Peru, while the sand rose of the dune field is shown in the inset. The barchans are crossing topographic rises, which suggests topography as cause for their asymmetric shape \citep{Bourke_2010}. Indeed, as discussed in Section \ref{sec:topography}, we found that when the barchans are {\em{upwind}} of the ridge's crest, the limb closest to the ridge appears extended, while the other limb elongates once the dune has crossed the ridge, i.e. when the dune is {\em{downwind}} of the ridge's crest (see Fig.~\ref{fig:topographic_raise}c). Although this behavior could explain the asymmetric shape of some of the dunes in Fig.~\ref{fig:topographic_raise}a (in particular see the simulation of Fig.~\ref{fig:topographic_raise}b and the more detailed discussion in Section \ref{sec:topography}), for many of the dunes in Peru other asymmetry causes cannot be excluded, in particular collisions \citep{Bourke_2010}.

\subsection{\label{sec:example_3}Third example: barchan asymmetry induced by collisions}

We see in Fig.~\ref{fig:barchan_field}c the third example of barchan asymmetry, which is caused by collisions. Many of the dunes in  Fig.~\ref{fig:barchan_field}c that are well separated from other dunes, and are thus not participating of a collision, have approximately the same shape, which is nearly symmetric and consistent with the wind regime indicated by the sand rose in the inset. Other dunes in the barchan corridor, especially those migrating close to other dunes, display different types of asymmetry. We see that several barchans have the upper limb extended, while others have the lower limb extended. Moreover, in some cases either limb is giving rise to a small barchan, thus indicating that dune {\em{calving}} \citep{Duran_et_al_2005,Genois_et_al_2013,Worman_et_al_2013} is a relevant mechanism for the generation of small dunes in the field. The simulations of Figs.~\ref{fig:barchan_field}a and \ref{fig:barchan_field}b, performed with strictly unidirectional wind regimes (wind blows from left to right), produce barchan asymmetry behavior with the characteristics found in the examples of Fig.~\ref{fig:barchan_field}c, thus indicating that complex (asymmetric) wind regimes are not a pre-requisite for the asymmetric dune shapes in the referred field.

\subsection{\label{sec:planetary_discussion}Application of the model to the research of extraterrestrial dunes --- a discussion}

Asymmetric barchan dunes occur in many dune fields on Mars \citep{Bourke_et_al_2004,Bourke_and_Goudie_2009,Bourke_2010}. It was shown by \cite{Bourke_2010} that the origin of a variety of Martian asymmetric barchans could be understood on the basis of the four main asymmetry causes tested in the present paper. Indeed, the insights of our simulations should be applicable for understanding the formation of seif dunes on Mars and also the influence of topography and collisions on complex dune patterns occurring within Martian barchan corridors. 

In Martian dune fields hosting asymmetric barchans with one limb elongating into a seif dune, Figs.~\ref{fig:asymmetric_wind2} and \ref{fig:asymmetric_wind3} could be used for estimating the divergence angle ${\theta}_{\mathrm{w}}$ and the ratio $r$ between the transport rates associated with the dominant sand-moving wind directions. We note that a diversity of exotic Martian dune shapes occurring on top of bedrock could be indeed simulated with the present model using obtuse bimodal wind regimes with $r = 1.0$ \citep{Parteli_and_Herrmann_2007,Parteli_et_al_2009}. On the other hand, due to the lower gravity of Mars the influence of topography on the shape of barchan dunes on the red planet should be correspondingly smaller than on Earth. For instance, we expect dunes crossing a sloping terrain on Mars to migrate laterally as they do on Earth, however the value of $k_q$ (cf.~Eq.~(\ref{eq:vT_vL})), which gives the ratio of transverse to longitudinal migration rates, should be smaller on Mars for a given terrain's slope. Furthermore, simulations of dune collisions such as the ones in Figs.~\ref{fig:dune_collision2} and \ref{fig:barchan_field} but using attributes of sediment and fluid valid for Martian sand systems could shed light on the origin of complex asymmetric patterns occurring on different locations (and thus environmental conditions) on Mars. However, in order to perform quantitative simulations, the model should be further improved by accounting for recently noted differences between saltation on Mars and Earth as well as for the recently gained insights into the physics of the threshold wind velocity required to sustain sediment transport on Mars \citep{Almeida_et_al_2008,Kok_2010}. Moreover, future simulations should incorporate a model for the saturation length ${\ell}_{\mathrm{s}}$ (cf.~Eq.~(\ref{eq:sand_flux})) that encodes the relevant mechanisms of flux saturation of sediment transport in extraterrestrial environments \citep{Paehtz_et_al_2013}. 

Finally, asymmetric barchans leading to seif dunes due to limb extension could be identified in Cassini images of the surface of Titan \citep{Radebaugh_et_al_2010}, thus indicating asymmetric bimodal wind regimes on this Saturn's moon. The formation of the equatorial linear dune fields of Titan has been indeed attributed to seasonally varying winds \citep{Lorenz_et_al_2006,Tokano_2010}. Thus, the modeling presented here could be used to investigate the ratios of transport rates and divergence angles of Titan's bimodal wind regimes at those locations hosting asymmetric barchans.

\section{\label{sec:conclusions}Conclusions}

In conclusion, we have investigated, by means of morphodynamic modeling of aeolian dunes, the respective roles of bimodal wind regimes, topography, influx asymmetry and dune collisions on the formation and evolution of asymmetric barchan dune shapes. The conclusions of our numerical simulations can be summarized as follows:
\begin{itemize}
\item {\em{Bimodal wind regimes}} --- the limb opposite to the secondary wind elongates into the resultant transport direction if the divergence angle of the bimodal wind is obtuse and the ratio ${\left[{Q_2T_{{\mathrm{w}}2}}\right]}/{\left[{Q_1T_{{\mathrm{w}}1}}\right]}$ --- where $Q_1$ ($Q_2$) and $T_{{\mathrm{w}}1}$ ($T_{{\mathrm{w}}2}$) stand for the bulk sand flux and the duration of the primary (secondary) wind, respectively --- exceeds $25\%$. These conditions are the same as the ones for oblique alignment of bedforms in dense sand beds \citep{Rubin_and_Hunter_1987}, the asymmetric barchan being the corresponding morphology under low amount of available sand.
\item {\em{Topography}} --- a barchan crossing a topographic rise can become asymmetric; the limb closest to the topographic rise elongates downwind. The migration velocity of a barchan that is on a tilted surface has a downhill component proportional to the tilting slope; the preferential limb extension is downhill rather than downwind.

\item {\em{Influx asymmetry}} --- the side of the barchan subjected to the larger influx increases in volume, whereas the opposite limb elongates downwind. The typical asymmetric morphology is a barchan with an elongated arm (the one subjected to lower influx). The asymmetry is of transient nature, since the dune migrates laterally towards the region of higher influx.

\item {\em{Dune collisions}} --- asymmetry can be triggered due to barchan collisions with lateral offset. Binary collisions with large offset can lead to extension of the limb resulting from the merging of the two limbs participating at the collision. Collisions between multiple barchans in a field can trigger more complex asymmetric patterns. Some patterns produced in these simulations involve the elongation of one limb of the upwind dune, as reported from previous observations \citep{Bourke_2010}.
\end{itemize}

In order to improve the quantitative assessment of asymmetric barchans, the model should be extended in order to account for secondary flow effects at the dune lee \citep{Tsoar_2001}, which may be relevant for the dynamics of limb elongation due to bimodal wind regimes or due to topographic ridges with sharp slopes \citep{Bourke_2010}. Indeed, the separation bubble model of Section \ref{sec:wind_field} does not account for the occurrence of three-dimensional flow structures within the zone of recirculating flow, which affect the dune shape at the lee and can contribute to the elongation of the limb. Realistic simulations of bimodal wind regimes should further account for multiple secondary wind trends in consistence with the complex sand roses of real dune fields \citep{Fryberger_and_Dean_1979}. Moreover, real winds can be very gusty, with changes in direction of 45 degrees or more taking place in most cases in less than 15 minutes, while in some instances the changes may be gradual or abrupt \citep{Elbelrhiti_et_al_2005,Duran_et_al_2011}. This effect of changes in wind directions should be considered in the future. Finally, we also note that there are examples of seif dunes formation from vegetated linear dunes that lost their vegetation due to human impact \citep{Tsoar_2001}. The formation of seif dunes in the presence of a dynamic vegetation cover remains to be investigated in the future.

The results of our calculations are potentially useful for inferring local wind regimes or spatial variations either in topography or in availability of mobile sediments in planetary dune fields where asymmetric barchans occur. 

\section*{Acknowledgments}
This work was supported in part by NASA MDAP Grant NNX10AQ35G, by FUNCAP, CAPES and CNPq (Brazilian agencies), by FUNCAP Grant 0011-00204.01.00/09, by Swiss National Foundation Grant NF 20021-116050/1 and ETH Grant ETH-10 09-2. We also thank the German Research Foundation (DFG) for funding through the Collaborative Research Initiative ``Additive Manufacturing'' (SFB814) and the Cluster of Excellence ``Engineering of Advanced Materials''.

\end{document}